\documentclass[final,5p,twocolumn,10pt]{elsarticle} 

\usepackage{lineno}
\usepackage[draft]{hyperref}
\usepackage{enumitem}
\modulolinenumbers[5]

\usepackage{amsmath}
\usepackage{amssymb}
\usepackage{stmaryrd}
\usepackage{amsthm}
\usepackage{amsfonts}
\usepackage{dsfont}
\usepackage{graphicx}
\usepackage{upgreek}
\usepackage{bm}
\usepackage{color}
\usepackage{float}
\usepackage{tikz-cd}
\usepackage{relsize}

\usepackage{mathtools}
\usepackage{algorithm,algpseudocode} 
\usepackage{todonotes}

\usepackage{subcaption}

\usepackage{tabu}
\usepackage{dblfloatfix} 

\biboptions{numbers,sort&compress}

\newcommand{\bzero}{{\mathbf{0}}}

\newcommand{\bx}{{\mathbf{x}}}

\newcommand{\bt}{{\mathbf{t}}}
\newcommand{\bk}{{\mathbf{k}}}
\newcommand{\bu}{{\mathbf{u}}}

\newcommand{\bff}{{\mathbf{f}}}

\newcommand{\sign}{\mathsf{sign}}

\newcommand{\indic}{{\mathbf{1}}}

\newcommand{\vol}{\mathsf{vol}}

\newcommand{\sweep}{\mathsf{sweep}}

\newcommand{\R}{{\mathds{R}}}

\newcommand{\SO}[1]{{\mathrm{SO}(#1)}}
\newcommand{\SE}[1]{{\mathrm{SE}(#1)}}

\newcommand{\conf}{{\mathsf{C}}}

\newcommand{\imf}{IMF}
\newcommand{\targ}{\text{target}}
\newcommand{\acc}{\text{acc}}
\newcommand{\con}{\text{con}}
\newcommand{\unc}{\text{unc}}

\theoremstyle{definition}

\newcommand{\eq}[1]{(\ref{#1})} 
\newcommand{\com}[1]{} 

\journal{Computer-Aided Design}

\begin{document}

\begin{frontmatter}

\title{Topology Optimization with Accessibility Constraint for Multi-Axis Machining}
 
\author{ Amir M. Mirzendehdel, Morad Behandish, and Saigopal Nelaturi}
\address{\rm Palo Alto Research Center (PARC), 3333 Coyote Hill Road, Palo Alto, California 94304  \vspace{-15pt}}

\begin{abstract}
	
In this paper, we present a topology optimization (TO) framework to enable
automated design of mechanical components while ensuring the result can be
manufactured using multi-axis machining. Although TO improves the part's
performance, the as-designed model is often geometrically too complex to be
machined and the as-manufactured model can significantly vary due to machining
constraints that are not accounted for during TO. In other words, many of the
optimized design features cannot be accessed by a machine tool without colliding
with the part (or fixtures). The subsequent post-processing to make the part
machinable with the given setup requires trial-and-error without guarantees on
preserving the optimized performance. Our proposed approach is based on the
well-established accessibility analysis formulation using convolutions in
configuration space that is extensively used in spatial planning and robotics.
We define an \emph{inaccessibility measure field} (IMF) over the design domain to
identify non-manufacturable features and quantify their contribution to
non-manufacturability. The IMF is used to penalize the sensitivity field of
performance objectives and constraints to prevent formation of inaccessible
regions. Unlike existing discrete formulations, our IMF provides a continuous
spatial field that is desirable for TO convergence. Our approach applies to
arbitrary geometric complexity of the part, tools, and fixtures, and is highly
parallelizable on multi-core architecture. We demonstrate the effectiveness of
our framework on benchmark and realistic examples in 2D and 3D. We also  show
that it is possible to directly construct manufacturing plans for the optimized
designs based on the accessibility information.

\end{abstract}


\end{frontmatter}


\section{Introduction} \label{sec_intro}

Recent advances in computation and manufacturing technologies have enabled
engineers to improve quality, increase productivity, and reduce cost by
automating various stages of design and production. However, in many cases the
discrepancies between as-design and as-manufactured models can result in
excessive trial-and-error cycles or even render the design completely
non-manufacturable. Incorporating manufacturing constraints early on during the
design stage is essential for successful automation of computational design
workflows.

Topology optimization (TO) \cite{Bendsoe2009topology,Sigmund2013topology} is a
computational tool for automated design that enables engineers across multiple
disciplines ranging from biomedical \cite{Challis2010prototypes} to automotive
\cite {Wang2004automobile} and aerospace \cite{Zhu2016topology} explore the
expansive design space of functional components. The interest in TO stems from
recent advances in computational capabilities, new materials, and manufacturing
technologies, where multi-functional components can be optimized with high
fidelity to generate complex ``organic'' shapes that reduce cost while improving
performance. Advances in additive manufacturing (AM) have enabled engineers to
produce complex geometries designed by TO. However, many industrial parts
require high precision and surface quality that, as of today, can only be
achieved by subtractive manufacturing (SM) technologies such as multi-axis
machining. Current advances in automated manufacturing technologies have also
enabled hybrid manufacturing (HM) processes that combine the complementary
capabilities of AM and SM to achieve customization, cost-effectiveness,
geometric complexity, precision, and surface quality for industrial
functionality
\cite{Zhu2013review,Lorenz2015review,Flynn2016hybrid,Merklein2016hybrid}.

\begin{figure*}[t!]
	\centering
	\includegraphics[width=\linewidth]{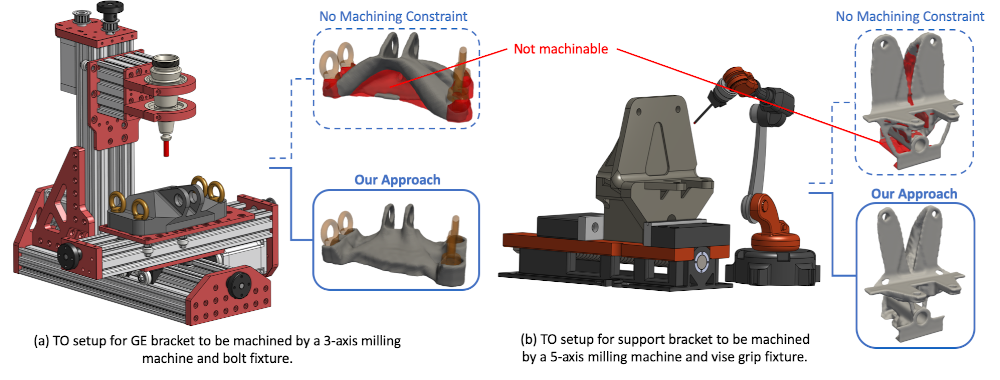}
	\caption{This article presents a generic approach to incorporate accessibility
	constraints for multi-axis machining into the design optimization process. The
	inputs are the shapes of initial design (e.g., the brackets shown above),
	stationary obstacles such as fixtures, moving objects including the entire tool
	assembly, and the available orientations. The output is an optimized design
	that is guaranteed to be accessible everywhere in the negative space of the
	part in at least one configuration and the set of all such configurations at
	the desired sampling resolution. The examples are demonstrative of arbitrary
	geometric complexity, and do not correspond to real industrial setups.}
\label{fig_TOsetups}
\end{figure*}

The focus of this paper is on developing a TO framework based on sound
mathematical concepts from spatial planning \cite{Lozano-Perez1983spatial} to
incorporate multi-axis machining constraints early on during the design stage.
The present work will substantially reduce the time and resources spent on
post-optimization trial-and-error by bridging the gap between design and a
widely used set of manufacturing processes. We will demonstrate the
effectiveness of our proposed framework by considering realistic examples for 3-
and 5-axis milling setups as shown in Fig. \ref{fig_TOsetups}. Our method relies
on a quantification of inaccessibility by any combination of translational and
rotational degrees of freedom (DOF) as a continuous spatial field (called IMF)
in 3D, which can be rapidly computed in parallel for intermediate designs of
arbitrary geometric and topological complexities generated by TO.

In the remainder of this section, we review recent advances in design for
manufacturing (DfM)---with a focus on incorporating manufacturability
considerations into TO---and accessibility analysis (Section \ref{sec_lit}). We
then present the contributions and outline of the paper (Section
\ref{sec_cont}).

\subsection{Related Work} \label{sec_lit}

The literature on TO with manufacturability constraints is broad, and our brief
overview of the most relevant works in this section is by no means
comprehensive. More detailed surveys can be found in
\cite{Zuo2006manufacturing,Sutradhar2017incorporating,Liu2016survey,Liu2018current}.

With the growing interest in AM processes, design for additive manufacturing
(DfAM) has emerged as an important paradigm to include AM constraints in early
stages of design. These constraints pertain to minimum feature size
\cite{Zhou2015minimum}, efficient use of support structure
\cite{Mirzendehdel2016support,Langelaar2016topology,Qian2017undercut},
anisotropic material properties
\cite{Mirzendehdel2018strength,Vantyghem2018compliance}, and post-processing
\cite{Liu2016survey,Langelaar2019integrated}.

For traditional manufacturing processes, DfM-based TO has mainly focused on
minimum feature size, casting, linear and axial symmetry, and wire-cut for 2.5D
structures. For casting and profile milling, the final design is constrained to
have no undercuts or cavities along a prescribed draw/mill direction. A number
of approaches have been used such as filtering of density fields
\cite{Zhou2002progress}, filtering of sensitivity fields
\cite{Wang2017structural}, projecting parametric design variables onto element
density spaces  \cite{Guest2012casting,Vatanabe2016topology}, and imposing
limits on the maximum of a virtual temperature field \cite{Li2018topology}.

For machining with few DOF (e.g., 2.5D profiling), the range of solutions for TO
can be restricted in terms of manufacturable ``features'' of shape. For
instance, a feature-based shape optimization was proposed in \cite{liu20153d} by
incorporating a feature-fitting algorithm into levelset TO. The feature-based TO
was extended to include a limited set of hybrid (combined AM/SM) manufacturing
constraints for 2.5D profiling \cite{liu2017topology} and overhang-free 3D
printing \cite{Liu2019topology}. Unfortunately, feature-based modeling presents
several ambiguities for process planning that prevent scalability with
increasing geometric complexity. Increasing the DOF adds to the complexity of
defining and identifying features for intermediate designs generated by TO.

Projection methods, on the other hand, are particularly suitable for designing
modular layouts as in structures constructed by joining (e.g., welding)
parameterized primitives such as bars or plates
\cite{Norato2015geometry,Zhang2016geometry}. A unified projection-based scheme
was developed in \cite{Vatanabe2016topology} to consider various constraints
including minimum member size, minimum hole size, symmetry, pattern repetition,
extrusion, turning, casting, forging, and rolling by applying different variable
mapping rules. Recently, a TO framework for 5-axis machining was proposed in
\cite{Langelaar2019topology}, where the sensitivity field is filtered by
accumulation of densities along tool insertion directions for a given set of
simple prismatic tool shapes, insertion depths, and directions. It demonstrates
that coupling morphological operations with projection methods is an effective
approach to enforce machining constraints in TO. However, the assumption that
the tool can approach the part only along straight line paths can over-constrain
the TO with complex tool shapes and motion DOF.

A common limitation of feature-based or projection-based approaches is that a
wide range of nontrivial designs with freeform features that can only be
fabricated by complex tool shapes and motions might be excluded from the design
space parameterization. Moreover, they do not rigorously formulate collision
avoidance between the stationary objects (i.e., workpiece and fixtures) and
moving objects (i.e., entire tool assembly). To realize the full potential and
flexibility of TO in generating non-parametric shapes, there is a need for a
more generic mathematical formulation of accessibility that 1) applies to
candidate designs of arbitrary shape, not necessarily modeled via feature
taxonomies or projected from predefined geometric building blocks; 2) provides a
continuous quantification of inaccessibility to guide gradient/sensitivity-based
TO; and 3) can be computed rapidly within a TO loop.

Accessibility and collision avoidance have been extensively studied in spatial
planning and robotics \cite{Lozano-Perez1983spatial,Latombe2012robot}. The
notion of a configuration space ($\conf-$space) of relative 6D motions, i.e.,
group of rigid-body translations and rotations, is introduced to abstract
collision predicates between two arbitrary 3D rigid bodies in relative motion to
a point membership query against a 6D pointset, referred to as the configuration
space obstacle ($\conf-$obstacle). Computing $\conf-$obstacles for arbitrary
shapes can be challenging, depending on the choice of representation. For 5-axis
machining applications, one of the six DOF for rotation around the tool axis is
deemed redundant and the problem is simplified by assuming simple shapes for the
rotating tool's closure such as a flat/ball-end cylinders, approximating
reachability by visibility, and so on
\cite{Woo1990visibility,Elber1994accessibility,Yin2002accessibility,
	Segall2014line,Kim2015precise,Ezair2018automatic,Sosin2019accessibility,
	Bo2019initialization}. However, the computations can be dramatically simplified
by sampling-based approximations \cite{Nelaturi2012rapid}. It has been shown
that collision measures can be obtained as convolutions of indicator functions
of the two bodies \cite{Lysenko2010group}, and computed rapidly via fast
Fourier transform (FFT) if these functions are sampled over uniform grids
(i.e., voxelization) \cite{Kavraki1995computation}. The convolution field
provides an implicit representation of the $\conf-$obstacle as its
$0-$superlevel set. Moreover, the field value provides a measure of collision
that quantifies inaccessibility and varies continuously with motion. In this
paper, we show how this 6D field can be projected back to 3D to be used as a
penalty function for TO.

In a recent article \cite{Mirzendehdel2019exploring}, we presented a formal
definition of inaccessibility measure that is well-suited for TO. The proposed
measure is agnostic to geometric complexities of part, tool assembly, and
surrounding fixtures, as well as the motion DOF. However, its effectiveness was
demonstrated only for simple 2D examples in \cite{Mirzendehdel2019exploring}
using a Pareto tracing levelset TO \cite{Suresh2010199}. In this paper, we
generalize the inaccessibility measure and extend its implementation to 3D for
density-based TO, and demonstrate it effectiveness by optimized non-parametric
designs with manufacturability guarantees for high-axis CNC machining with
arbitrary fixture and tool assembly shapes and cutting orientations. To the best
of our knowledge, none of the published TO frameworks provide matching extensive
flexibilities.

\subsection{A Note on Constraint Classification}

The accessibility constraint is typically classified as a `set constraint' and
is expressed in the language of set containment, interference, affine
transformations, and Boolean operations rather than the more commonly used
real-valued functions that appear in (in)equality constraints for TO (e.g.,
bounds on stress or deformation) and other conventional constrained optimization
problem formulations. To properly formulate the optimization problem by
simultaneously considering set constraints for manufacturing and (in)equality
constraints for performance, we rely on the classification scheme presented in
\cite{Mirzendehdel2019exploring}. We classify the constraints as global, local,
and strictly local (i.e., pointwise) constraints.

Global constraints are evaluated in terms of a real-valued property over the
entire design, such as compliance or p-norm of stress. Local constraints are
evaluated at each point of the design but are also dependent on membership of
other points in the design. Although they can also be interpreted as global
constraints, there are advantages to local formulation (i.e., as a field in 3D)
for TO. Accessibility constraint falls under this category, as collisions
occurring anywhere in the design, possibly far from a given query point, affects
deciding and quantifying the (in)accessibility of that point. Finally, strictly
local (or pointwise) constraints can be evaluated at every point of the design
without any knowledge of the membership of other points in the design.
Accessibility constraint becomes pointwise only when the motion is independent
of the global shape, e.g., if the collisions can only occur between the tool and
fixtures, regardless of the workpiece's evolving shape. See
\cite{Mirzendehdel2019exploring} (Section 4.4) for an example with 2.5D
wire-/laser-cutting of a sheet material. Pointwise constraints directly lead to
the definition of a point membership classification (PMC) for a maximal pointset
that represents the entire feasible design subspace of the constraint, hence is
used to `prune' the design space prior to optimization. In this paper, we are
interested in the more complex case of high-axis motions in which this
assumption is invalid and the part's own shape plays an important role in
quantifying (in)accessibility.

Global constraints are widely used in TO and are typically expressed as a
differentiable functional, for which a continuous sensitivity field can be
computed at every point in the 3D design domain. We showed in
\cite{Mirzendehdel2019exploring} that local constraints can be directly
incorporated into the sensitivity field as a penalty. However, if the penalty
function is not continuous, the resulting discontinuities in the ``augmented''
sensitivity field can adversely affect the convergence of TO to useful designs.
Although the set constraint for accessibility can be easily converted to an
(in)equality constraint in terms of discontinuous indicator functions (e.g., 1
for inaccessible points and 0 for accessible points) or a continuous field in 6D
(e.g., the convolution), it is not trivial to define it as a continuous field in
the 3D space for motions with arbitrary rotational DOF in multi-axis machining.
For small tools and purely translational motions, an approximation to IMF can be
obtained by establishing a simple one-to-one correspondence between the
translation $\conf-$space and 3D design domain by selecting a representative
point on the cutter. Every point in the design domain is assigned the value of
convolution corresponding to the collision measure for the translation that
takes the representative point to the query point. In
\cite{Mirzendehdel2019exploring}, we demonstrated that the resulting field makes
for an effective penalty for filtering the sensitivity field for producing
manufacturable designs in simple 2D examples with small 2D tools at a few
orientations. In this paper, we present a generic definition of IMF that is
usable for arbitrary shapes in 2D and 3D and motions including rotations.

\subsection{Contributions \& Outline} \label{sec_cont}

This paper presents a TO methodology to design high-performance lightweight
structures while also guaranteeing {\it accessibility} of every point on the
design's negative space for a given {\it collection} of cutting tool assemblies
and fixtures (arbitrary shapes in 3D) and available motions including
translations and rotations. The rotations can represent either a limited set of
fixturing orientations for 3-axis milling, or a discrete sample of orientations
that a 5-axis CNC system can reach.

Our approach does not impose any artificial limitations on geometric complexity
of part, tools, and fixtures. It enables efficient and effective design space
exploration to discover nontrivial geometries and topologies that are optimized
for the specific machining capabilities of a machine shop. The constrained
optimization is with regards to not only performance criteria (e.g., stiffness
or strength) but also manufacturability, as opposed to the conventional approach
that disregards the latter and postpones manufacturability concerns to
downstream post-processing.

More specifically, the contributions of this paper are:
\begin{enumerate}
	\item introducing a rigorous mathematical formulation of a continuous
	`inaccessibility measure field' (\imf) in 3D design domain to modify the
	sensitivity field for TO;
	\item formulating a TO framework that incorporates accessibility constraints
	for multi-axis machining based on realistic cutting tool assemblies and
	fixtures and arbitrary motions including translations and rotations;
	\item developing efficient and flexible implementation of the IMF that enables
	balancing numerical accuracy against available time budget for calling within
	the TO alongside the finite element analysis (FEA);
	\item implementing IMF on multi-core CPU/GPU for massive parallelization;
	\item applying the accessibility constraints for multi-axis machining to
	density-based TO; and
	\item demonstrating the effectiveness of our method by solving multiple
	benchmark and realistic examples in 2D and 3D on 2-, 3-, and 5-axis CNC
	machines.
\end{enumerate}
\section{Proposed Method} \label{sec_method}

In this section, we will first discuss our analytic approach to accessibility
analysis and introduce a continuous field to measure inaccessibility of a part
with respect to a collection of tools and fixtures at a discrete set of
fixturing orientations (Section \ref{sec_IMF}). Next, we extend the TO
formulation for incorporating multi-axis machining constraint into the
density-based TO framework (Section \ref{sec_TO}).

\subsection {Quantifying Multi-Axis Inaccessibility} \label{sec_IMF}

TO typically starts with an initial design $\Omega := \Omega_0 \subset \R^3$
(called the \emph{design domain}) and incrementally updates the design $\Omega
\subseteq \Omega_0$ such that it remains within the design domain while
minimizing the specified objective function and satisfying the specified
constraints. These constraints may include performance criteria (e.g., stiffness
or strength), evaluated by a physics solver such as FEA, as well as kinematic
constraints (e.g., machine tool accessibility), which require spatial analysis.
While the former is represented by (in)equality constraints in terms of
real-valued functions, the latter is most naturally expressed using a
set-theoretic language in terms of containment, interference, affine
transformations, and Boolean operations. Here, we will present an analytic
approach to convert the latter to (in)equality form to be used alongside the
former.

On a multi-axis CNC machine, one deals with 6D rigid motions $(R, \bt) \in
\SE{3}$, which are conceptualized as points in the {\it configuration space}
($\conf-$space) $\SE{3}$, i.e., a pair formed by an special orthogonal (SO)
automorphism of $\R^3$ (i.e., a 3D rotation) $R \in \SO{3}$ and a vector (i.e.,
a 3D translation) $\bt \in \R^3$. For 2- or 3-axis milling, the rotation
component is fixed at a finite set of fixturing orientations, while the tool is
swept along a continuum set of 2D or 3D translations. For 5-axis milling, there
are two additional DOF for rotations, since the rotation around the tool axis is
redundant.%
\footnote{The rapidly turning tool profile is typically modeled by its
	axisymmetric closure around the spindle axis (e.g., a flat/ball-end cylinder)
	rather than explicitly accounting for the rotation in $\conf-$space.}
The discrete fixturing orientations or continuum rotation DOF can be
parameterized in a number of different ways, e.g., $3 \times 3$ orthogonal
matrices, axis-angle pairs, unit quaternions, and Euler angles, or
can be combined with the translational element to form unified representations
such as $4 \times 4$ homogeneous matrices, dual quaternions, screws, etc.
Each have their own pros and cons, which are
well-understood. Our formulation is not restricted to a specific
parameterization of $\SO{3}$.

In practice, the workspace of the CNC machine is a bounded subset of $\SE{3}$
which is digitized into a discrete set (i.e., finite sample) in accordance with
the machine's precision and required algorithmic accuracy.

For spatial planning, the obstacles $O := (\Omega \cup F)$ consist of the
part/workpiece $\Omega \subset \R^3$ (i.e., evolving portion via TO) and the
fixtures $F \subset \R^3$ (i.e., fixed portion), both of which are 3D pointsets
represented in the same global coordinate frame. The tool assembly $T = (H \cup
K)$ consists of the tool holder $H \subset \R^3$ (i.e., inactive portion)  and
the cutter $K \subset \R^3$ (i.e., active portion) represented in the same local
coordinate frame, which is transformed by the {\it relative} rigid
transformation $(R, \bt) \in \SE{3}$ with respect to the global coordinate frame
of stationary obstacles.%
\footnote{In reality, both workpiece and tool assembly may move. Since
	accessibility depends only on relative motion, we can assume the former to be
	stationary without loss of generality.}

Assuming that the raw stock is the same as the design domain $\Omega_0$, the
accessibility constraint can be formulated as follows: for every point on the
part's exterior within the raw stock (i.e., the \emph{negative space}) $(\Omega_0 -
\Omega)$, there must exist a transformation $(R, \bt) \in \SE{3}$ that brings at
least one point on the cutter (hereon called a \emph{sharp point}) $\bk \in K$ in
contact with the query point, {\it without incurring a volumetric collision}
between the objects in relative motion:
\begin{align}
	&\forall \bx \in (\Omega_0 - \Omega): \quad \exists~(R, \bt) \in \SE{3}
	~\text{and}~ \exists ~ \bk \in K \nonumber \\
	&\text{s.t.}~ \bx = (R, \bt)\bk = R \bk + \bt ~\text{and}~ O \cap^\ast
	(R, \bt)T = \emptyset, \label{eq_access_def}
\end{align}
where the asterisk in $\cap^\ast$ stands for regularization after intersection
\cite{Tilove1980closure}, i.e., touching only at the boundaries does not count
as a collision, thus would not violate the above condition. $(R, \bt)T = RT +
\bt$ stands for the transformed tool assembly (rotation before translation).

\subsubsection{Morphological Definition of Accessibility} \label{sec_morph}

The accessibility is commonly formulated in terms of the configuration space
obstacle ($\conf-$obstacle) of relative transformations. The $\conf-$obstacle is
defined as the set of all transformations that result in a collision, violating
\eq{eq_access_def}:
\begin{equation}
	\mathcal{O} := \big\{ (R, \bt) \in \SE{3} ~|~ O \cap^\ast (R, \bt)T \neq
	\emptyset \big\}. \quad \label{eq_conf}
\end{equation}
The accessible region $A \subseteq \Omega_0$, defined by the set of all points
in the design domain that satisfy \eq{eq_access_def}, can be computed by {\it
	sweeping} (i.e., morphological dilation) of the cutter along the maximal
collision-free motion. The latter is obtained as the complement of
$\conf-$obstacle in the $\conf-$space (i.e, the `free space') $\mathcal{O}^c
= \SE{3} - \mathcal{O}$, hence:
\begin{align}
	A(O, T, K) :=&~ \Omega_0 \cap ~\sweep(\mathcal{O}^c, K) \\
	=& ~ \Omega_0 \cap \!\!\! \bigcup_{(R, \bt) \in \mathcal{O}^c}
	\!\!\! (RK + \bt). \label{eq_sweep}
\end{align}
Both sweeps and $\conf-$obstacles can be expressed in terms of Minkowski
products in $\conf-$space, and, in turn, as unions of the more familiar
Minkowski sums in $\R^3$ if the rotations are factored out as follows
\cite{Lysenko2010group}:
\begin{equation}
	A(O, T, K)  = \Omega_0 \cap \!\!\!\bigcup_{R \in \SO{3}} \!\!\!
	(O \oplus (-RT))^c \oplus (RK), \label{eq_A}
\end{equation}
in which $\oplus, \ominus, (\cdot)^c$ are the Minkwoski sum, Minkowski
difference, and set complement, respectively. For a given orientation $R \in
\SO{3}$, the first sum $D := O \oplus (-RT)$ is a translational ``slice'' of the
$\conf-$obstacle, whose complement $D^c$ is the collection of all collision-free
translations (i.e., a slice of $O^c$ for a fixed rotation). The second sum $D^c
\oplus (RK)$ represents the accessible region for the same orientation, obtained
by sweeping the rotated cutter $RK$ along the maximal collision-free translation
$D^c$. The inaccessible region $B \subseteq \Omega_0$ is the set of points in
the raw stock that do not belong in $A$:
\begin{equation}
	B(O, T, K) := \Omega_0 - A(O, T, K). \label{eq_B}
\end{equation}

\begin{figure*} [t!]
	\centering
	\includegraphics[width=0.95\linewidth]{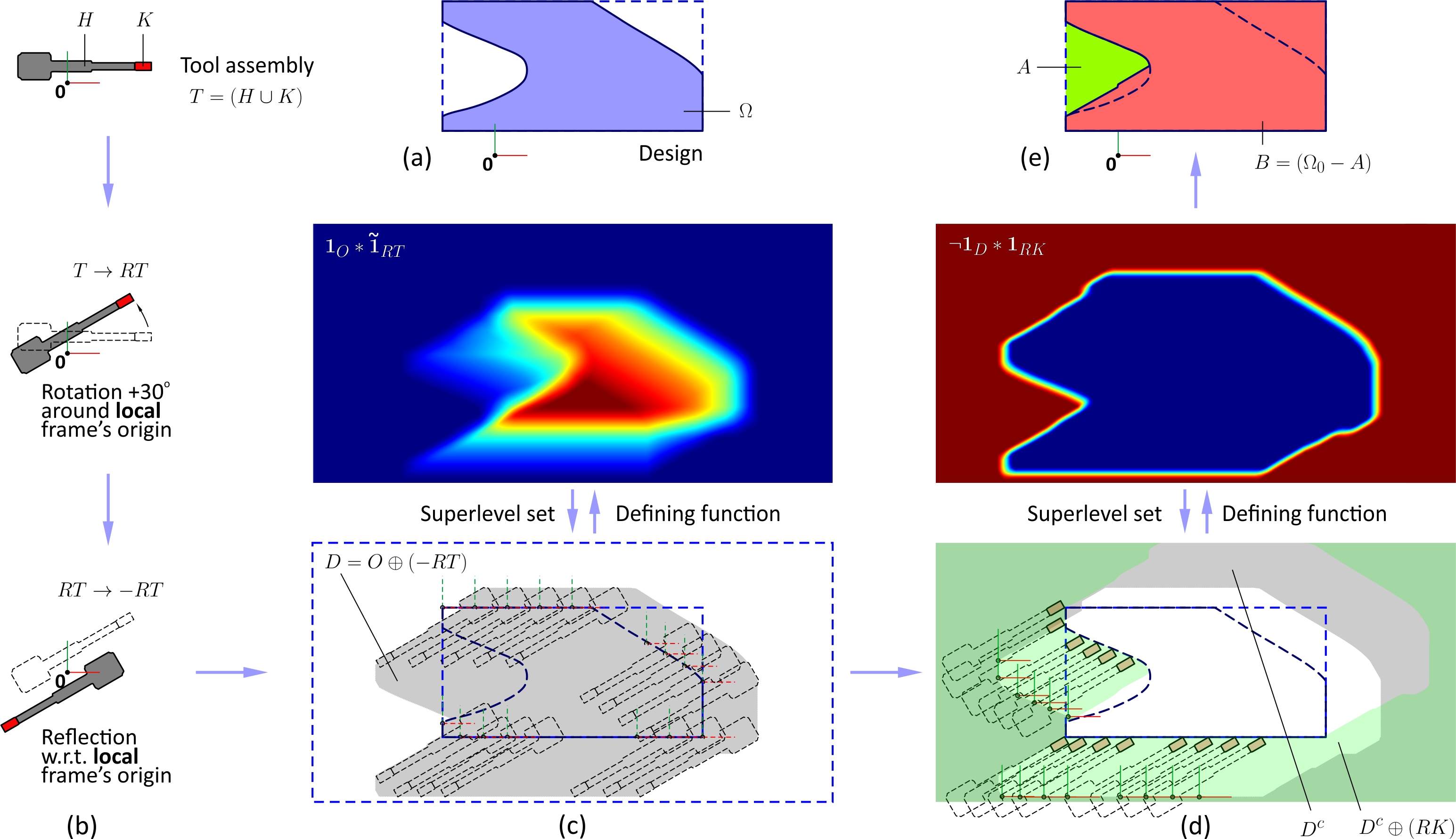}
	\caption{Consider (a) a 2D part $O = \Omega$ (here $F = \emptyset$) and (b) a
	2D tool assembly $T = (H \cup K)$. The tool is re-oriented as $T \to RT$ for a
	given $R \in \SO{2}$ and reflected as $RT \to -RT$. (c) The Minkowski sum $D =
	O \oplus (-RT)$ gives the set of colliding translations for the fixed rotation,
	which can be obtained as $0-$superlevel set of a convolution $(\indic_D \ast
	\tilde{\indic}_{RT})$. (d) The Minkowski sum $D^c \oplus (RK)$ gives the
	accessible region as the sweep of the cutter along collision-free translations,
	which can be obtained as $0-$superlevel set of a convolution $(\neg \indic_D
	\ast {\indic}_{RK})$. The decomposition into accessible $A$ and inaccessible
	$B$ regions is shown in (e).} \label{fig_morph}
\end{figure*}

To convert the global set-theoretic definition of accessibility to a local
(in)equality constraint, we use the correspondence between Minkowski and
convolution algebras for explicit and implicit morphology, respectively
\cite{Lysenko2010group}. The indicator function of any pointset $X \subseteq
\R^3$ is a binary-valued field denoted by $\indic_X: \R^3 \to \{0, 1\}$ defined
as:
\begin{equation}
	\indic_X (\bx) = \left\{
	\begin{array}{ll}
		1 & \text{if}~ \bx \in X; \\
		0 & \text{otherwise.}
	\end{array}
	\right. \quad \text{for any}~ X \subseteq \R^3.
\end{equation}
Under fairly general regularity conditions,%
\footnote{The correspondence is only valid if the participating sets are
	homogeneously 3D, e.g., the free space has no singularities, which is sufficient
	for our purposes. See \cite{Behandish2017analytic} (Section 3.4) for details.}
we have:
\begin{align}
	\indic_{D} (\bt) &= \sign \circ (~~ \indic_{O} \! \ast \tilde{\indic}_{RT}~\!)
	(\bt), \label{eq_conv_D} \\
	\indic_{A} (\bx) &= \sign \circ (\neg \indic_{D} \ast \indic_{RK}) (\bx),
	\label{eq_conv_A}
\end{align}
where $\ast$ stands for the convolution operator defined for integrable fields
over $\R^3$, and $\indic_{D^c}(\bt) = \neg \indic_D(\bt)$ where $\neg$ stands
for logical negation. We introduce the notation $\tilde{\indic}_X(\bx) =
\indic_{-X}(\bx) = \indic_X(-\bx)$ for reflection with respect to the origin,
which is needed to match how convolutions are defined to attain their useful
properties. The sign function is defined as $\sign(x) = x/|x|$ if $x \neq 0$ and
zero otherwise, and is required for algebraic closure---to convert the
real-valued convolution to a binary-valued indicator function before passing it
to the next convolution. See \ref{app_conv} for an explanation of the above
identities. Fig. \ref{fig_morph} illustrates both explicit and implicit
operations in 2D.

While the indicator functions are useful for accessibility analysis as a post-TO
test, we need a spatial field to penalize {\it inaccessibility} of different
points within the candidate design $\Omega \subseteq \Omega_0$ to prevent the TO
from violating accessibility at every iteration.

\subsubsection{Inaccessibility Measure as Convolution}

The no-collision condition in \eq{eq_access_def} can also be expressed in terms
of the {\it measure} of intersection:
\begin{equation}
	O \cap^\ast (R, \bt)T = \emptyset ~\rightleftharpoons~ \vol\big[O \cap
	(R, \bt)T \big] = 0, \label{eq_access_meas}
\end{equation}
where $\vol[\cdot]$ stands for volume (i.e., Lebesgue $3-$measure) of a 3D
pointset. This measure can be computed as an inner product of indicator
functions, i.e., integration of their product over $\R^3$. For objects in
relative motion, the translational component results in a shift of function
argument, turning the inner product into a convolution:
\begin{equation}
	\vol\big[O \cap (R, \bt)T \big] = \langle \indic_O, \indic_{(R, \bt)T} \rangle
	= (\indic_{O} \ast \tilde{\indic}_{RT}) (\bt), \label{eq_ident}
\end{equation}
which also appeared on the right-hand side of \eq{eq_conv_D}.

At a first glance, the convolution field appears like an ideal candidate for
penalization in TO: a continuous field over $\R^3$ that measures
inaccessibility. At a closer look, however, the domain of this function is the
translational $\conf-$space, which is a different ``type'' than the design
domain. The former is a space of 3D displacement {\it vectors} (i.e., position
{\it differences}) while the latter is of 3D {\it points} (i.e., positions). The
convolution function measures the inaccessibility for a hypothetical
displacement of $\bt \in \R^3$ that has nothing to do with any point $\bx \in
\Omega_0$. The function shifts with different choices of origin for the local
coordinate system in which the tool assembly is described.%
%

To properly ``register'' the shifted field with the design domain, we must
select the origin at the sharp points so that the convolution $(\indic_{O} \ast
\tilde{\indic}_{RT})(\bt)$ evaluated at the translation $\bt \in \R^3$ returns
the collision measure for shifting the sharp point from the origin $\bzero$ to
$\bx = (R, \bt)\bzero = R \bzero + \bt = \bt$. Since we have more than one
option for the sharp point, each one provides an independent candidate for the
origin to register the two spaces by shifting the convolution.

\subsubsection{Inaccessibility for A Single Tool Assembly} \label{sec_single}

\begin{figure*} [t!]
	\centering
	\includegraphics[width=\linewidth]{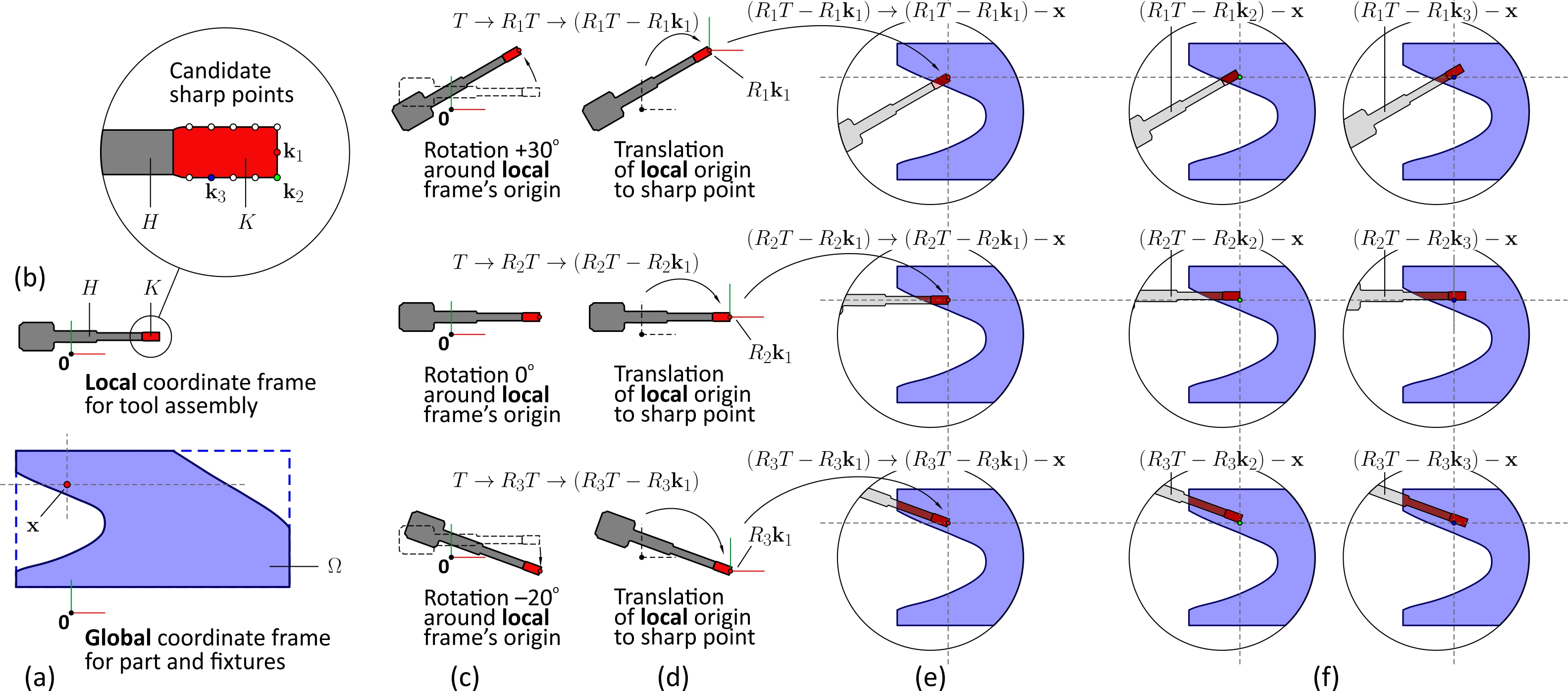}
	\caption{Consider (a) a 2D part and (b) a 2D tool assembly. At a given query
		point $\bx \in \Omega_0$, the IMF is computed by looking at different rotations
		$R_1, R_2, R_3, \ldots \in \Theta$ of the tool assembly, 3 of which are shown
		in (c). For each oriented tool $T \to R_i T$, the origin is shifted to
		different sharp points $\bk_1, \bk_2, \bk_3, \ldots  \in K$; e.g., for the
		(rotated) sharp point $R_i \bk_j \in R_i K$, the tool is translated as $R_i T
		\to (R_iT - R_i \bk_j)$ shown in (d). The new origin is brought in contact with
		the query point, hence $(R_iT - R_i \bk_j) \to (R_iT - R_i \bk_j) -\bx$ shown
		in (e). This is repeated for all candidate sharp points (f) and the IMF is
		computed as the minimum over all rotations and sharp points.} \label{fig_mins}
\end{figure*}

We define the \emph{inaccessibility measure field} (IMF) over the 3D design domain
$f_\text{IMF}: \R^3 \to \R$ for each given tool assembly $T = (H \cup K)$ as the
pointwise {\it minimum} of shifted convolutions for different choices of sharp
points and available orientations $\Theta \subseteq \SO{3}$ (which depends on
$T$):
\begin{equation}
	f_\text{IMF}(\bx; O, T, K) := \min_{R \in \Theta} \min_{\bk \in K} ~\vol
	\big[ O \cap (R, \bx) (T - \bk) \big]. \label{eq_imf_1}
\end{equation}
There are two independent transformations in effect:
\begin{itemize}
	\item The shift $T \to (T - \bk)$ in \eq{eq_imf_1} is to try different ways to
	register the translation space with the design domain, by changing the local
	coordinate system to bring different sharp points to the origin.
	\item The rotation $(T - \bk) \to (RT - R\bk)$ followed by translation $(RT -
	R\bk) \to (RT - R\bk) + \bx$ bring the candidate sharp point (new origin) to
	the query point $\bx \in \Omega_0$.
\end{itemize}
The same effect can be obtained by querying the convolution in \eq{eq_ident} at
$\bt := (\bx - R\bk)$ so that the rigid transformation $(R, \bt)$ brings the
sharp point in contact with the query point: $(R, \bt)\bk = R\bk + \bt = R\bk +
(\bx - R\bk) = \bx$, as expected. The IMF is thus computed as follows:
\begin{equation}
	f_\text{IMF}(\bx; O, T, K) = \min_{R \in \Theta} \min_{\bk \in K}
	~(\indic_{O} \ast \tilde{\indic}_{RT})(\bx - R\bk). \label{eq_imf_2}
\end{equation}
Each transformed convolution measures the collision for an attempt to remove the
query point $\bx \in \Omega_0$ in the candidate orientation $R \in \Theta$ with
the sharp point $\bk \in K$. The inaccessibility of the query point is
determined by the orientation and sharp point that result in the {\it best case}
scenario, i.e., the least collision volume.

Figure \ref{fig_mins} illustrates the idea behind \eq{eq_imf_2} for simple 2D
shapes with a few candidate orientations and sharp points. Note that the
collision measure accounts for both: 1) penetrations of the cutter into the
local neighborhood of the part (often referred to as ``gouging''
\cite{Kim2015precise}) or fixtures, which leads to over-cutting; and 2) global
interferences that may occur elsewhere along the tool holder.

The IMF can be used to classify the design domain into disjoint subsets $A :=
f_\text{IMF}^{-1}(0)$, and $B = \Omega_0 - A$:
\begin{align*}
	A(O, T, K) := \big\{ \bx \in \Omega_0 ~|~ f_{\text{IMF}}(\bx; O, T, K) = 0
	\big\}, \\
	B(O, T, K) := \big\{ \bx \in \Omega_0 ~|~ f_{\text{IMF}}(\bx; O, T, K) > 0
	\big\}.
\end{align*}
which are the same as the pointsets defined in \eq{eq_sweep} and \eq{eq_B} if
$\Theta := \SO{3}$, under quite general conditions. A query point is accessible
iff its IMF is zero, i.e., there exists one or more tool orientations and sharp
points with which the query point can be touched without incurring a collision.

\begin{figure} [t!]
	\centering
	\includegraphics[width=\linewidth]{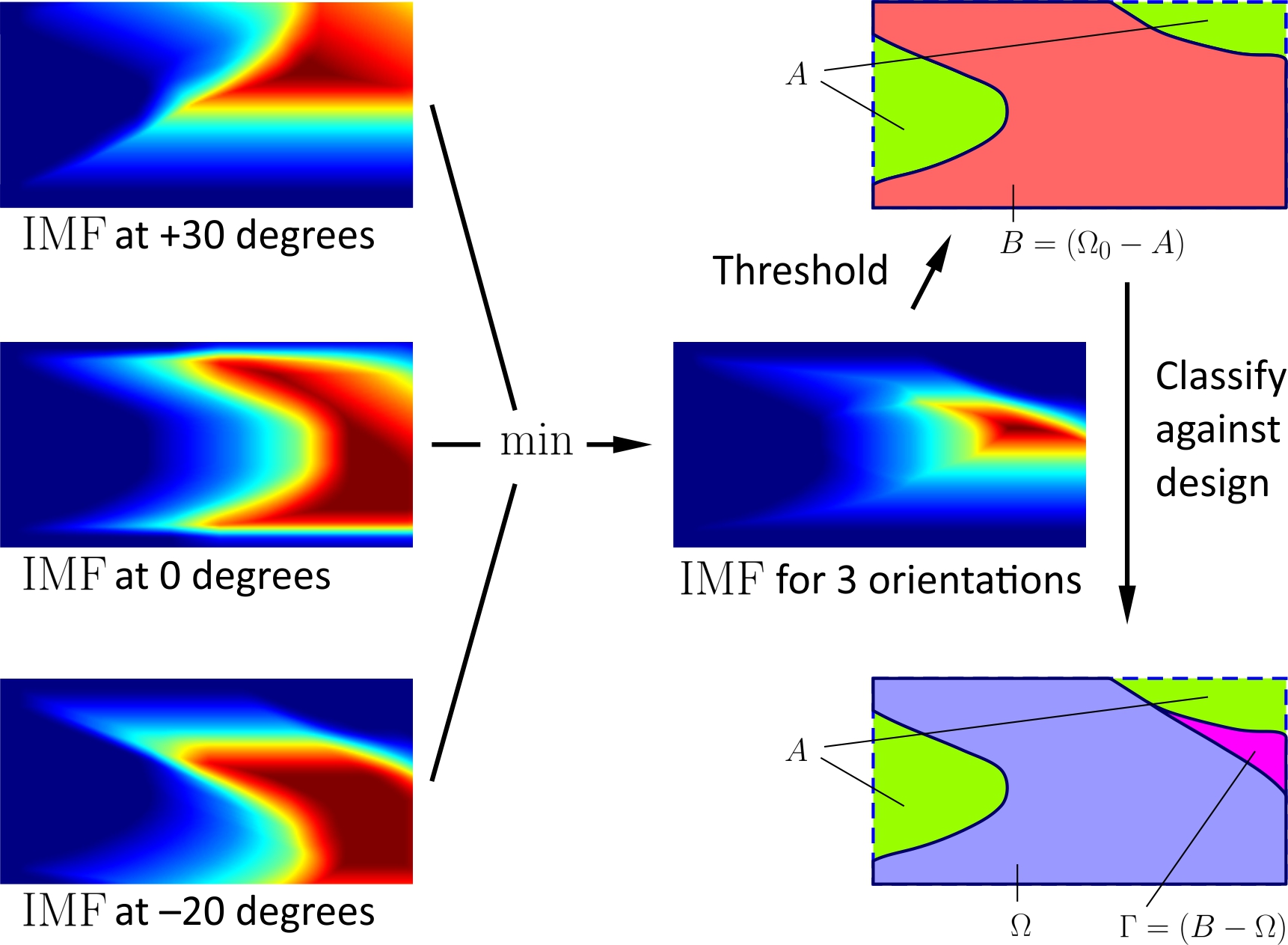}
	\caption{For every candidate orientation in Fig. \ref{fig_mins} (c), the
		minimum of convolution over all sharp points is computed. The overall IMF is
		obtained as another minimum over all orientations, whose zero/nonzero values
		determine accessible/inaccessible regions.} \label{fig_fields}
\end{figure}

Note that every point inside the design itself is inaccessible, i.e., $\Omega
\subseteq B$ thus $(\Omega \cap A) = \emptyset$. Hence, the inaccessible region
can be further decomposed into two disjoint subsets, the part $\Omega$ and
$\Gamma := (B - \Omega)$, to which we refer as the `secluded region'. The latter
is the set of all points in the negative space $(\Omega_0 - \Omega)$ of the
part/workpiece that are inaccessible, i.e., points in the raw stock that cannot
be machined at any orientation with the given tool using the specified options
for sharp points. Figure \ref{fig_fields} illustrates this three-way
decomposition for a 2D example.

For parts designed for machining, TO must resist generating secluded regions,
including nucleation of internal voids inside the part, which are often seen in
TO parts meant for AM.

\subsubsection{Inaccessibility for Multiple Tool Assemblies} 

Given $n_T \geq 1$ available tool assemblies $T_i = (H_i \cup K_i)$ for $1 \leq
i \leq n_T$, we compute their combined IMF by applying another minimum operation
over different tools to identify the tool(s) with the smallest volumetric
interference at available orientations and sharp points:
\begin{equation}
	f_\text{IMF}(\bx; O) := \min_{1 \leq i \leq n_T} f_\text{IMF}(\bx; O, T_i, K_i)
	\label{eq_imf_3}
\end{equation}
in which $f_\text{IMF}(\bx; O, T_i, K_i)$ are computed from \eq{eq_imf_2}. Once
again, we can decompose the design domain into accessible and inaccessible
regions, respectively, with respect to all available tool assemblies:
\begin{align}
	A(O) &:= \!\!\! \bigcup_{1 \leq i \leq n_T} \!\!\! A(O, T_i, K_i),
	\label{eq_A_multi} \\
	B(O) &:= \!\!\! \bigcap_{1 \leq i \leq n_T} \!\!\! B(O, T_i, K_i).
	\label{eq_B_multi}
\end{align}
in which $A(O, T_i, K_i)$ and $B(O, T_i, K_i)$ were obtained earlier. The
secluded region with respect to all tools is the subset of inaccessible regions
that lies outside the design:
\begin{equation}
	\Gamma(O) := B(O) \cap (\Omega_0 - \Omega) = B(O) - \Omega. \label{eq_gamma}
\end{equation}

\subsubsection{Algorithm to Support Density-Based TO} \label{sec_disc_alg}

In density-based TO, one deals with a continuous density function
${\rho}^{}_\Omega: \Omega \to [0, 1]$ to represent intermediate designs, rather
than indicator functions. While we can use a threshold $0 < \tau < 1$ (e.g.,
$\tau := 0.5$) to define the indicator functions as $\indic_\Omega (\bx) := 1$
iff $\rho^{}_\Omega (\bx) > \tau$ for use in \eq{eq_imf_2}, our experience shows
that direct use of the density function works better to provide additional
smoothing:
\begin{equation}
	f_\text{IMF}(\bx; \rho^{}_O, T, K) := \min_{R \in \Theta} \min_{\bk \in K}
	~(\rho^{}_{O} \ast \tilde{\indic}_{RT})(\bx - R\bk). \label{eq_imf_4}
\end{equation}
The function $\rho^{}_O: \Omega_0 \to [0, 1]$ can be obtained as:
$\rho^{}_O(\bx) := \rho^{}_\Omega(\bx) + \indic_F(\bx)$, in which
$\rho^{}_\Omega(\bx)$ is obtained directly from TO. The combined IMF for all
tool assemblies $f_\text{IMF}(\bx; \rho^{}_O)$ is computed as:
\begin{equation}
	f_\text{IMF}(\bx; \rho^{}_O) := \min_{1 \leq i \leq n_T}
	f_\text{IMF}(\bx; \rho^{}_O, T_i, K_i) \label{eq_imf_5}
\end{equation}

The IMF is in units of volume. To use it alongside other constraints in TO
(e.g., stiffness and strength) we first normalize it by the global maximum to
obtain $\bar{f}_\text{IMF}(\bx; \rho^{}_O)$.

Moreover, to deal with inaccuracies due to discretization and intermediate
densities in TO, we allow for some relaxation, aiming for a small allowance of
$0 < \lambda \ll 1$ (e.g., $\lambda := 0.01$) for nonzero inaccessibility. In
other words, we only penalize the points that have (normalized) inaccessibility
value of greater than the threshold $\lambda$.

For the sake of generality, we assume that a given tool assembly $T_i$ comes
with a given set of rotations $\Theta_i \subset \SO{3}$ available for orienting
that tool. For 3-axis milling, the set of available rotations is finite,
corresponding to different fixturing configurations. For 5-axis milling, we can
assume a continuum set of rotations, which can be sampled for computational
purposes.

In practice, the shape of fixtures $F$ hence $O = (\Omega \cup F)$ can change
every time the part is rotated and re-clamped for 3-axis milling at a different
orientation. For clarity, we do not consider multiple fixtures in this paper,
although accounting for their changing shapes comes at no extra cost as long as the fixturing setup is given a priori.

Algorithm \ref{alg_imf2} describes the subroutine that computes the IMF for a
given candidate design and a specified set of tool assemblies, orientations, and
fixtures for use within TO iterations. For every tool assembly represented by a
3D array $[\indic_{T_i}]$ of binary values (i.e., voxels) and a given
orientation $R \in \Theta_i$ represented by a unit quaternion, the 3D array
$[\tilde{\indic}_{RT_i}]$ is obtained by rotation and re-sampling on the uniform
grid, followed by a reflection with respect to the origin.%
\footnote{The reflection is conveniently implemented as a conjugation in the
	frequency domain after FFT (Hermitian symmetry).}
The discretized convolution is represented as a 3D array $[\rho^{}_{O} \ast
\tilde{\indic}_{RT_i}]$ and is computed by two forward FFTs on $[\rho^{}_O]$ and
$[\tilde{\indic}_{RT_i}]$, a pointwise multiplication in the frequency domain,
and an inverse FFT back to the physical domain (of translations). This is
repeated for all tools and orientations to obtain the minimums in \eq{eq_imf_4}.
See \ref{app_comp} for an analysis of time and space complexity.

\begin{algorithm} [ht!]
	\caption{Compute $[\bar{f}^{}_{\text{IMF}}],[\indic_{\Gamma(O)}]$.}
	\begin{algorithmic}
		\Procedure{IMF}{$[\rho^{}_{\Omega}], [\indic_F], [\indic_{H_i}],
			[\indic_{K_i}], \{\Theta_i\}; \lambda, n_T$}
		\State Define $[\rho^{}_{O}] \gets [\rho^{}_{\Omega}] + [\indic_{F}]$
		\Comment{Implicit union}
		\State Initialize $[f_\text{IMF}] \gets 0$ \Comment{IMF for all the tools}
		\For{$i \gets 1$ to $n_T$}
		\State Define $[\indic_{T_i}] \gets [\indic_{H_i}] + [\indic_{K_i}]$
		\Comment{Implicit union}
		\State Initialize $[\gamma_i] \gets [0]$ 
		\Comment{IMF for the $i^\text{th}$ tool}
		\ForAll{$R \in \Theta_i$}
		\State $[\indic_{RT_i}] \gets \Call{Rotate}{[\indic_{T_i}], R}$
		\Comment{Re-sampling}
		\State $[\tilde{\indic}_{RT_i}] \gets \Call{Reflect}{[\indic_{RT_i}]}$
		\State $[g_{i}] \gets \Call{Convolve}{[\rho^{}_O], [\tilde{\indic}_{R T_i}]}$
		\Comment{FFT-based}
		\ForAll{$\bk \in \Call{Support}{[\indic_{K_i}]}$}
		\State $[h_{i}] \gets \Call{Translate}{[g_{i}], -R \bk}$
		\State $[\gamma_i] \gets \min([\gamma_i], [h_{i}])$
		\Comment{Over sharp points}
		\EndFor
		\EndFor
		\State $[f_\text{IMF}] \gets \min([f_\text{IMF}], [\gamma_i])$
		\Comment{Over all tools}
		\EndFor
		\State $[\bar{f}_\text{IMF}] \gets [\bar{f}_\text{IMF}] /
		\Call{Max}{[\bar{f}_\text{IMF}]}$ \Comment{Normalization}
		\State $[\indic_{B(O)}] \gets [\bar{f}_\text{IMF} > \lambda]$
		\State $[\bar{f}_\text{IMF}] \gets [\bar{f}_\text{IMF}] \cdot [\indic_{B(O)}]$
		\Comment{Add allowance}
		\State $[\indic_{\Gamma(O)}] \gets [\indic_{\Omega_0}] \cdot [\neg
		\indic_\Omega] \cdot [\indic_{B(O)}]$ \Comment{Implicit intersect}
		\State\Return{($[\bar{f}_\text{IMF}]$, $[\indic_{\Gamma(O)}]$)}
		\EndProcedure 
	\end{algorithmic} \label{alg_imf2}
\end{algorithm}

Our formulation in terms of sharp points allows immense flexibility for
balancing accuracy against computation time. As the cutter's boundary is sampled
more densely, the IMF can only decrease in value due to the minimum operation in
\eq{eq_imf_2}, and the set of secluded points $\Gamma(O)$ grows in size. This
comes at a small cost of more queries on the convolution. Importantly, coarser
sampling of the cutter can only {\it over-estimate} the exact IMF, leading to a
{\it conservative} approximation of inaccessibility. As more sharp points are
sampled on the cutter, more candidate designs are deemed machinable by carving
out their negative spaces via the same sample points. Omitting other sharp
points can over-constrain the TO by ``false positives'' in collision detection,
i.e., obtaining $f_{\text{IMF}}(\bx; \Omega) > 0$ while the true value is zero;
however, the approximation never violates the exact form of accessibility
constraint.

It is worthwhile noting that our model of inaccessibility does not distinguish
between different modes of collision such as local over-cutting (i.e., gouging)
and global interferences with part/fixturess. Our initial TO experiments showed
that defining different collision measures and using unequal weights when
penalizing them in TO show no notable improvement in the TO results. In fact,
imbalanced weighting can result in tolerating one type of collision in favor of
another, resulting in undesirable design artifacts.

\subsection {Machining-Constrained Topology Optimization} \label{sec_TO}

Based on the accessibility analysis discussed in Section \ref{sec_IMF}, we
formulate the TO problem as:
\begin{subequations} \label{eq_TOproblem}
	\begin{align}
		\mathop{\text{Minimize}}\limits_{\Omega \subseteq \Omega_0} \quad &
		\varphi(\Omega), \label{eq_TOproblem_a} \\
		\text{such that} \quad 
		& [\textbf{K}_\Omega][\textbf{u}_\Omega]=[\bff], \label{eq_TOproblem_b} \\
		& V_\Omega \leq V_{\targ}, \label{eq_TOproblem_c}\\ 
		& V_{\Gamma(O)} = 0, \label{eq_TOproblem_d} 
	\end{align}
\end{subequations}
where $\varphi(\Omega) \in \R$ is the value of objective function for a given
design $\Omega \subseteq \Omega_0$. $[\bff]$, $[\bu_\Omega]$, and
$[\textbf{K}_\Omega]$ are (discretized) external force vector, displacement
vector, and stiffness matrix, respectively, for FEA. $V_\Omega := \vol[\Omega]$
represents the design volume and $V_{\targ} > 0$ is the volume budget. The
accessibility constraint for machining is imposed via \eq{eq_TOproblem_d} by
asserting that the secluded $V_{\Gamma(O)} := \vol[\Gamma(O)]$ (i.e., volume of
inaccessible regions in the negative space) must be zero. In practice, we impose
a small nonzero upper-bound $\sim 1\%$ of part's volume) to provide relaxation
against discretization errors.
This initial formulation of accessibility as a `global' constraint makes it
difficult to incorporate into TO, as computing a local gradient/sensitivity for
the inaccessible volume with respect to design variables is theoretically
challenging, due to inherent discontinuities in collision detection, and
computationally prohibitive. However, we showed in
\cite{Mirzendehdel2019exploring} that by formulating the inaccessibility as a
`local' constraint $\bar{f}_\text{IMF}(\bx) \leq \lambda$ for all $\bx \in
(\Omega_0 - \Omega)$, the continuous IMF can be directly augmented into the
sensitivity field to filter out the inaccessible regions of the design domain.

Putting aside the accessibility constraint in \eq{eq_TOproblem_d} for the
moment, the more familiar constrained optimization problem of
\eq{eq_TOproblem_a} through \eq{eq_TOproblem_c} can be expressed as minimization
of the following Lagrangian:
\begin{equation}
	\mathcal{L}_{\Omega}:= \varphi(\Omega)
	+ ~\lambda_1 (\dfrac{V_\Omega}{V_{\targ}} - 1)
	+ [\lambda_2]^\mathrm{T} \Big([\textbf{K}_\Omega][\bu_\Omega] - [\bff]\Big).
	\label{eq_Lag}
\end{equation}
Using the prime symbol $(\cdot)'$ to represent the generic (i.e.,
representation-agnostic) differentiation of a function with respect to $\Omega$,
we obtain (via chain rule):
\begin{align}
	\mathcal{L}_\Omega' &= \varphi'(\Omega) + \lambda_1
	\dfrac{{V}_\Omega'}{V_{\targ}} + [\lambda_2]^\mathrm{T}
	\Big([\textbf{K}_\Omega][\bu_\Omega]\Big)',\\
	& = \Big([ \dfrac{\partial \varphi}{\partial \bu} ]+ [\lambda_2]^\mathrm{T}
	[\textbf{K}_\Omega]\Big)[\bu_\Omega'] \nonumber \\
	& +\lambda_1 \dfrac{{V}_\Omega'}{V_{\targ}} + [\lambda_2]^\mathrm{T}
	[\textbf{K}_\Omega'][\bu_\Omega]. \label{eq_chain_rule}
\end{align}
Since computing $[\bu_\Omega']$ requires solving \eq{eq_TOproblem_d} as many
times as the number of design variables and is computationally prohibitive,
$[\lambda_2]$ is chosen as the solution to the adjoint problem
\cite{Bendsoe2009topology} which reduces \eq{eq_chain_rule} to:
\begin{align}
	&\mathcal{L}_\Omega' = \lambda_1 \dfrac{{V}_\Omega'}{V_{\targ}}  +
	[\lambda_2]^\mathrm{T} [\textbf{K}_\Omega'][\bu_\Omega], \label{eq_Lag_prime}\\
	&\text{if} ~ [\lambda_2] := -[\textbf{K}_\Omega]^{-1}[\dfrac{\partial
	\varphi}{\partial \bu}]. \nonumber
\end{align}
When the objective function is the design's compliance under the applied load;
namely, $\varphi(\Omega) := [\bff]^\mathrm{T}[\bu_\Omega]$, we obtain
$[\lambda_2] = -[\bu_\Omega]$. This dramatically simplified the problem as the
compliance is self-adjoint, i.e., there is no need for solving an additional
adjoint problem unlike the case with other objective functions (e.g., stress).

To incorporate the accessibility constraint for multi-axis machining, we modify
the sensitivity field $	\mathcal{S}_\Omega$ as follows:
\begin{equation}
	\mathcal{S}_\Omega := (1-w_{\acc}) ~ \bar{\mathcal{S}}_\varphi + w_{\acc} ~
	\bar{\mathcal{S}}_{\text{IMF}}, \label{eq_filtered_sens}
\end{equation}
where $0 \le w_{\acc} < 1$ is the filtering weight for accessibility, and can be either a constant or adaptively updated based on the secluded volume $V_{\Gamma(O)}$.
$\bar{\mathcal{S}}_\varphi$ is the normalized sensitivity field with respect to the
objective function, i.e., only the second term
$[\lambda_2]^\mathrm{T}[\textbf{K}_\Omega'][\bu_\Omega]$ on the right hand side of \eq{eq_Lag_prime},
noting that the volume constraint is satisfied with the optimality criteria iteration \cite{sigmund200199}. $\bar{\mathcal{S}}_{\text{IMF}}$ is the
normalized accessibility filter defined in terms of the normalized IMF as:
\begin{equation}
	\bar{\mathcal{S}}_{\text{IMF}} (\bx) :=
	\begin{dcases}
	\bar{f}_\text{IMF}(\bx; \rho^{}_O) &  \text{if} ~ \bx \in \Omega, \\
	1 & \text{if} ~ \bx \in \Gamma(O), \\
	0 & \text{otherwise}.
	\end{dcases} \label{eq_accFilter}
\end{equation}	
in which $O = (\Omega \cup F)$ and $\rho^{}_{O}(\bx) = \rho^{}_\Omega(\bx) +
\indic_F(\bx)$ represent the design and fixtures, explicitly and implicitly.
$\bar{f}_\text{IMF}(\bx; \rho^{}_O)$ is obtained from \eq{eq_imf_5} (after
normalization by its maximum) and $\Gamma(O) = B(O) - \Omega = B(O) \cap
\Omega^c$ represents the secluded regions, i.e., inaccessible portion of the
negative space, define in \eq{eq_gamma}. Since we are working with implicit
representations, the first condition $\bx \in \Omega$ in \eq{eq_accFilter} is
computed by $\indic_\Omega(\bx) = 1$, i.e., $\rho^{}_\Omega(\bx) > \tau$. The
second condition $\bx \in \Gamma(O)$ is computed as a conjunction of 
$\bar{f}_\text{IMF} > \lambda$ (for $\bx \in B(O)$) and $\rho^{}_\Omega(\bx)
\leq \tau$ (for $\bx \in \Omega^c$).

The expressions in \eq{eq_TOproblem} through \eq{eq_accFilter} are general and
representation-agnostic, and can be used in both density-based and levelset TO.
To generate the results of this paper, we use the method of solid isotropic
material with penalization (SIMP). The implicit design representation
$\rho^{}_\Omega: \Omega_0 \to [0, 1]$ used in the definition of IMF is obtained
as the projection of another field $\xi^{}_\Omega: \Omega_0 \to [0, 1]$
(smoother density field for design exploration) whose discretization
$[\xi^{}_\Omega]$ is used as SIMP design variables. We use the following
Heaviside projection \cite{Andreassen2011efficient}:
\begin{equation}
	\rho^{}_\Omega (\bx) = 1 - e^{-\beta \xi^{}_\Omega(\bx)}+ \xi^{}_\Omega(\bx)
	e^{-\beta}, \label{eq_Heaviside}
\end{equation}
We use $\beta := 2$ for 2D and $\beta := 8$ for 3D examples of Section
\ref{sec_results}. Algorithm \ref{alg_TO} describes the
accessibility-constrained TO using SIMP. The sensitivity field is augmented
using the IMF output of Algorithm \ref{alg_imf2}.

\begin{algorithm} [ht!]
	\caption{TO with multi-axis machining constraint.}
	\begin{algorithmic}
		\Procedure{TO}{$[\bff],V_{\targ}, [\indic_{H_i}],[\indic_{K_i}],\{\Theta_i\};\lambda,n_T,\beta,\delta,l$}
		\State Initialize $[\xi_\Omega] \gets [V_\targ /
			\Call{Integral}{[\indic_{\Omega_0}]}]$
		\State Initialize $\Delta \gets \infty$
		\State Initialize $iter \gets 0$
		\While {$\Delta > \delta$ \textbf{and} $iter < l$ }
		\State $[\rho^{}_\Omega] \gets \Call{Heaviside}{[\xi_\Omega], \beta}$
			\Comment{Projection}
		\State $[\bu] \gets \Call{\text{FEA}}{[\rho^{}_\Omega], [\bff]}$
			\Comment{Solve FEA}
		\State $\varphi(\Omega) \gets \Call{\text{Evaluate}}{[\rho^{}_\Omega],
			[\bu]}$ \Comment{Obj. func.}
		\State $[\bar{\mathcal{S}}_\varphi] \gets
			\Call{\text{Gradient}}{[\rho^{}_\Omega], [\bu],
			\varphi(\Omega)}$\Comment{Sensitivity}
		\State $[\rho^{}_O] \gets [\rho^{}_\Omega] + [\indic_F]$
			\Comment{Implicit union}
		\State $([\bar{f}_\text{IMF}], [\indic_{\Gamma(O)}]) \gets
			\Call{IMF}{[\rho^{}_{\Omega}], [\indic_F], \ldots; \lambda, n_T}$
		\State \Comment{Call Algorithm \ref{alg_imf2} with obvious arguments}
		\State $[\bar{\mathcal{S}}_\text{IMF}] \gets
			\Call{Penalize}{\bar{f}_\text{IMF}, [\indic_{\Gamma(O)}]}$
			\Comment{Eq. \eq{eq_accFilter}}
		\State $[\mathcal{S}] \gets (1-w_{\acc}) [\bar{\mathcal{S}}_\varphi]
			+ w_{\acc} [\bar{\mathcal{S}}_{\text{IMF}}]$
			\Comment{Eq. \eq{eq_filtered_sens}}
		\State $[\xi_\Omega^{\text{new}}] \gets
			\Call{\text{Update}}{[\xi_\Omega], [\mathcal{S}]} $ 
			\Comment{SIMP filtering}
		\State $\Delta \gets \Call{Integrate}{[\xi_\Omega^{\text{new}}]
			- [\xi_\Omega]}$ \Comment{Vol. diff.}
		\State $iter \gets iter + 1$ \Comment{Iter.  counter}
		\State $[\xi_\Omega] \gets [\xi_\Omega^{\text{new}}]$
			\Comment{For next iteration}
		\EndWhile
		\State\Return{$[\xi_\Omega]$}
		\EndProcedure 
	\end{algorithmic} \label{alg_TO}
\end{algorithm}
\section{Results} \label{sec_results}

In this section, we will present benchmark and realistic examples in 2D and 3D.
All results are generated using a SIMP implementation, where the optimality
criteria method was used to update the density field.

All examples are run on a desktop machine with
\textsf{Intel}$^{\small{\textregistered}}$
\textsf{Core}$^{\small{\texttrademark}}$i7-7820X CPU with 8 processors running
at 4.5 GHz, 32 GB of host memory, and an
\textsf{NVIDIA}$^{\small{\textregistered}}$
\textsf{GeForce}$^{\small{\textregistered}}$ GTX 1080 GPU with 2,560
\textsf{CUDA} cores and 8 GB of device memory.

\subsection{Benchmark Example: Cantilever Beam}

\begin{figure} [ht!]
	\centering
	\includegraphics[width=\linewidth]{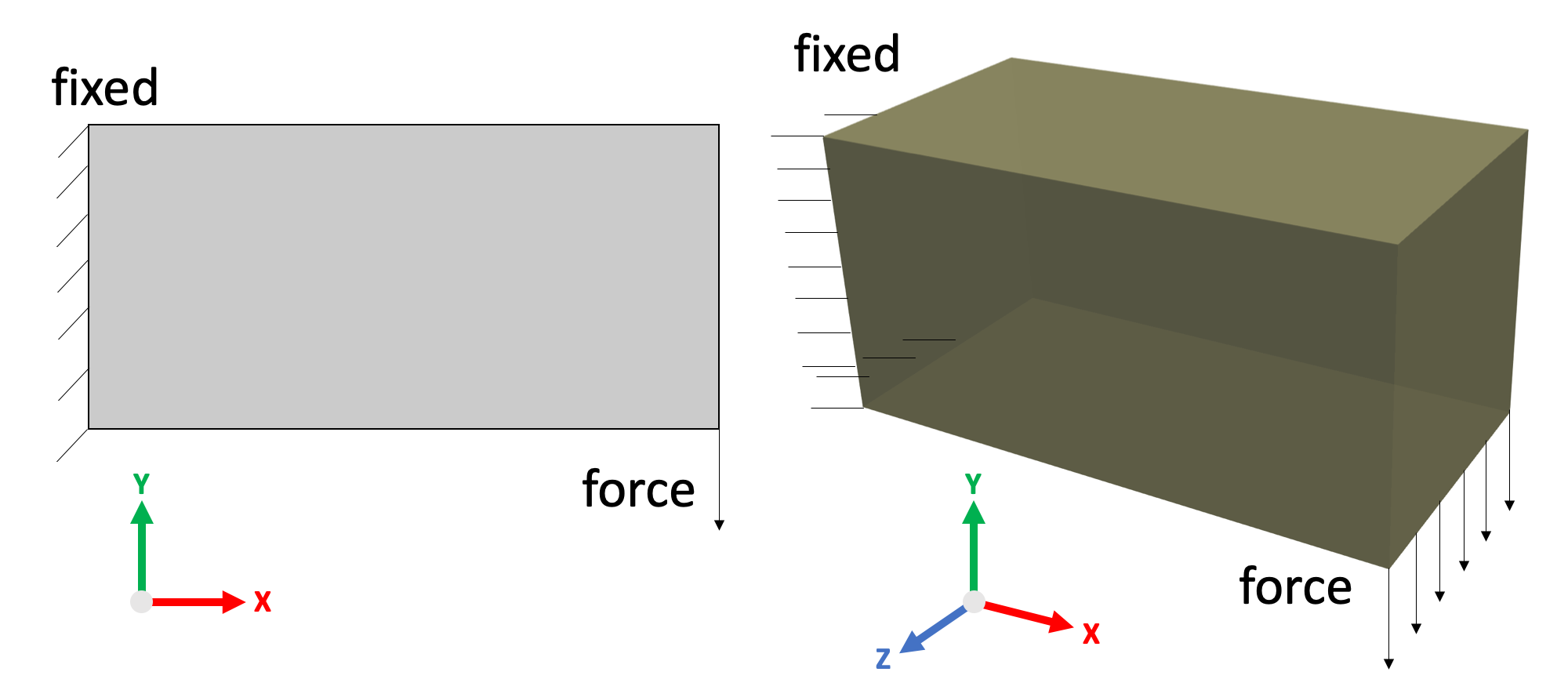}
	\caption{Cantilever beam boundary conditions in 2D and 3D.}
	\label{fig_beamLoading}
\end{figure}

First, we consider a simple cantilever beam example in 2D and 3D. The loading
conditions are shown in Fig. \ref{fig_beamLoading}. We use material properties
of Stainless Steel with Young's modulus of $E = 270$ GPa and Poisson ratio of
$\nu=0.3$. In each example, we solve both accessibility-constrained and
unconstrained TO and report the nonzero secluded volume for the latter
$V_{\Gamma_\unc} > 0$ whose prevention comes at the cost of an increase in
compliance results ($\varphi_\con > \varphi_\unc$).

\begin{figure} [ht!]
	\centering
	\includegraphics[width=\linewidth]{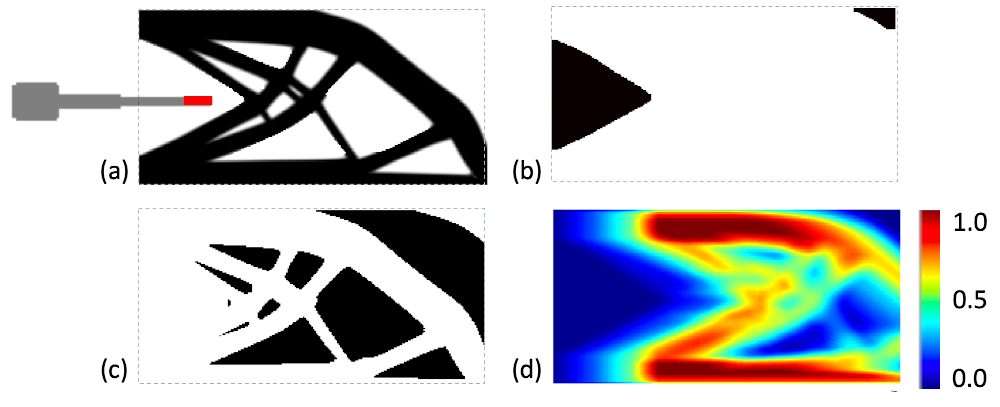}
	\caption{Accessibility-unconstrained TO for cantilever beam in 2D: (a)
		optimized cantilever beam at volume fraction of $0.5$ and the oriented end-mill
		tool, (b) set of collision-free translations of cutter, (c) secluded regions,
		and (d) normalized IMF.} \label{fig_mfgAnalysis2D}
\end{figure}

In 2D, we set the volume fraction to 0.5 and optimize the design for minimal
compliance with and without accessibility constraints for machining at
resolution 256$\times$128. The accessibility constraint is defined using two
cutting tool assemblies of nontrivial shapes, one with a thinner and another
with a thicker cutting edge. Fig. \ref{fig_mfgAnalysis2D} illustrates the
accessibility-unconstrained optimized cantilever beam at 0.5 volume fraction and
accessibility analysis based on the cutting tool with thin cutter with $(+1,0)$
tool direction.

\begin{figure} [t!]
	\begin{subfigure}[t]{\linewidth}
		\centering
		\includegraphics[width=\linewidth]{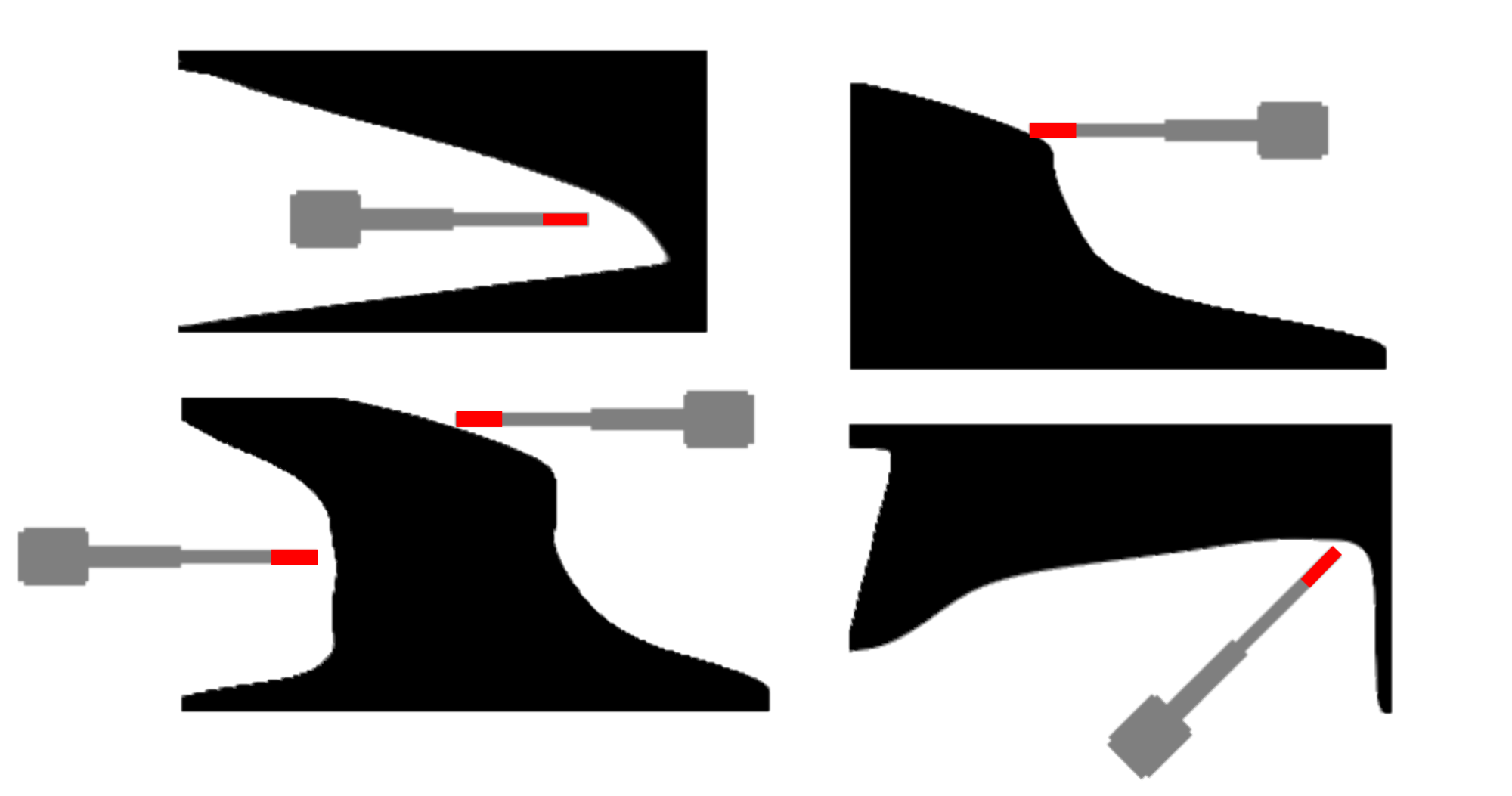}
		\caption{Thin cutter}
	\end{subfigure}%
	\\
	\begin{subfigure}[t]{\linewidth}
		\centering
		\includegraphics[width=\linewidth]{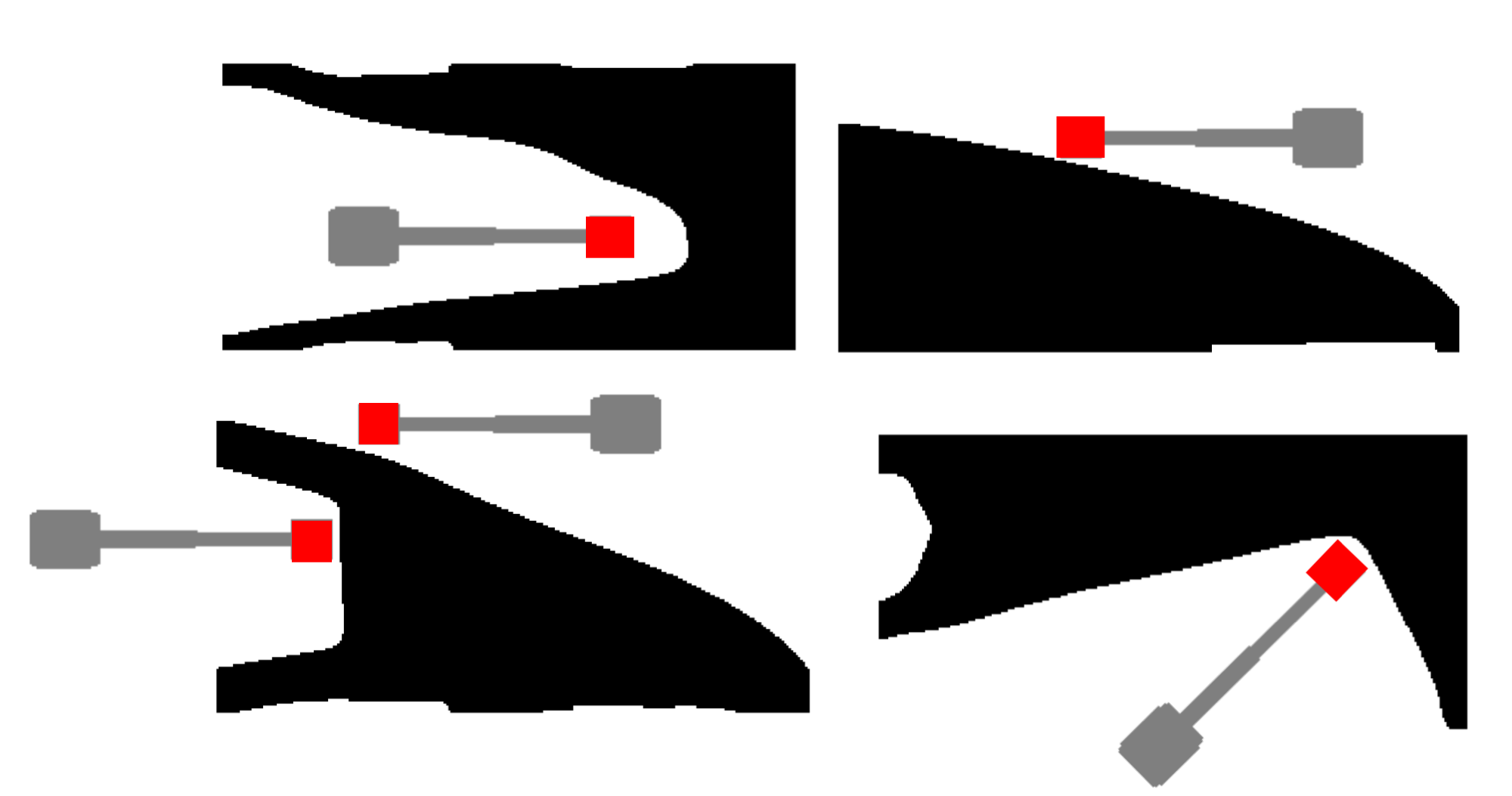}
		\caption{Thick cutter}
	\end{subfigure}%
	\caption{Benchmark cantilever beam example in 2D at a volume fraction of $0.5$.
		The accessibility constraint is imposed at 8 different configurations with two
		endmill cutter profiles.} \label{fig_benchmark2D}
\end{figure}

At first, let us consider only a few orientations per tool, one-at-a-time, for
which the tool configuration and the optimized designs are depicted in Fig.
\ref{fig_benchmark2D} (b, c). 

\begin{figure} [t!]
	\centering
	\includegraphics[width=\linewidth]{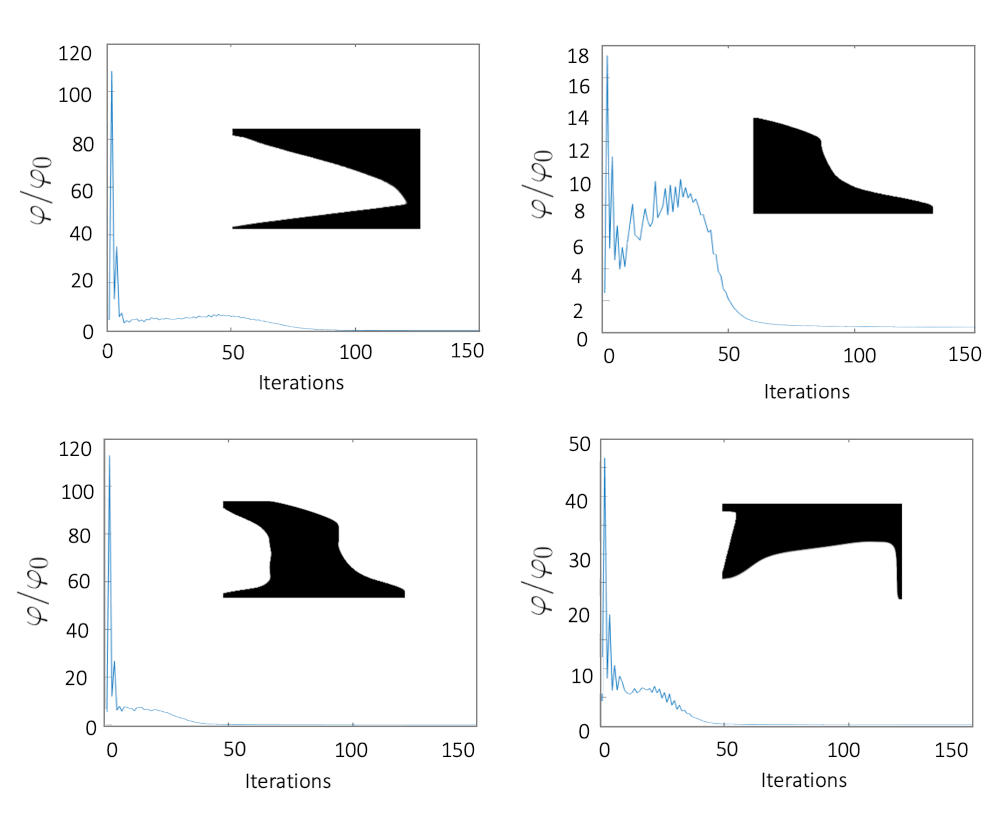}
	\caption{Convergence plots for 2D benchmark examples with thin cutter. $\varphi$ and $\varphi_0$ are complaince and complaince of the initial design with uniform density of 0.5.}   \label{fig_convergencePlots}
\end{figure}

\begin{figure} [t!]
	\centering
	\begin{subfigure}[t]{0.9\linewidth}
		\centering
		\includegraphics[width=\linewidth]{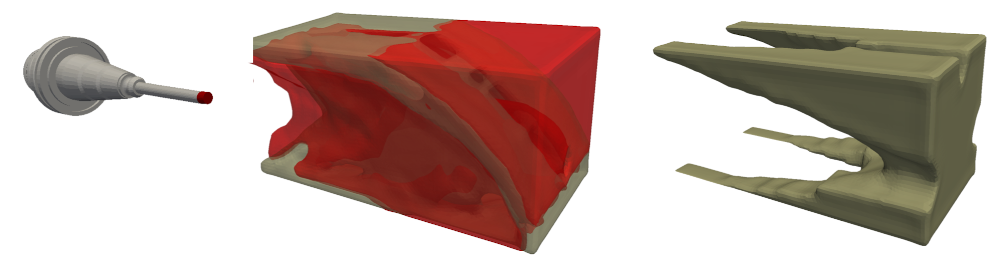}
		\caption{Tool orientation: $(+1,0,0)$}
	\end{subfigure}%
	\\
	\begin{subfigure}[t]{0.99\linewidth}
		\centering
		\includegraphics[width=\linewidth]{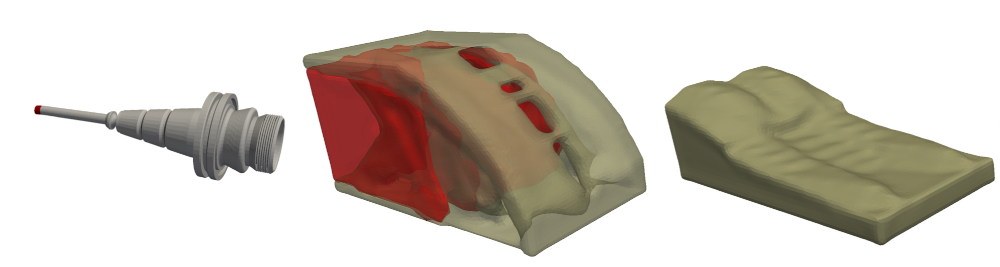}
		\caption{Tool orientation: $(-1,0,0)$}
	\end{subfigure}%
	\\
	\begin{subfigure}[t]{0.99\linewidth}
		\centering
		\includegraphics[width=\linewidth]{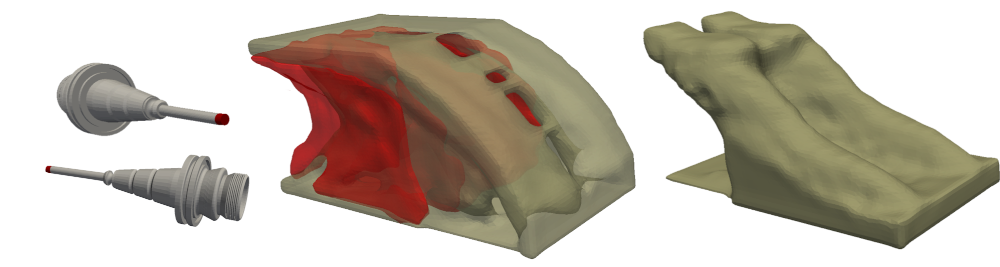}
		\caption{(Tool orientation: $(\pm1,0,0)$}
	\end{subfigure}%
	\\
	\begin{subfigure}[t]{0.99\linewidth}
		\centering
		\includegraphics[width=\linewidth]{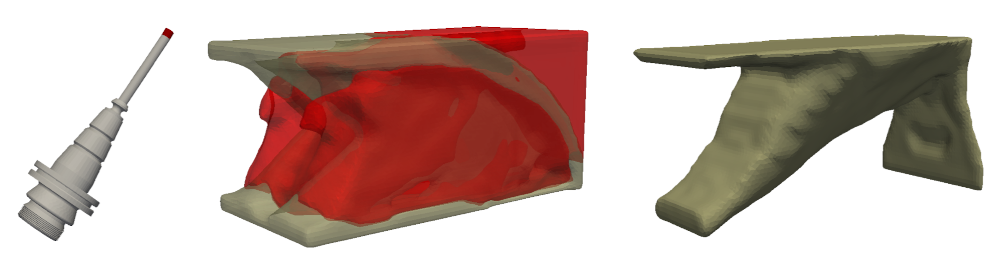}
		\caption{Tool orientation: $(+1,+1,0)$}
	\end{subfigure}%
	\caption{Benchmark cantilever beam example in 3D at $0.3$ volume fraction,
		using different tool orientation combinations. Secluded regions are shown in red.} \label{fig_benchmark3D}
\end{figure}

Observe that imposing accessibility constraints prevents nucleation of interior
holes in the optimized design. Further, the shape of cutter can alter the
accessible region at each iteration and IMF, subsequently changing the final
design. Table \ref{tab_beam2Dresults} summarizes the results for compliance and
secluded volume in 2D for the tool in Fig. \ref{fig_benchmark2D} (a) for
accessibility-unconstrained and constrained cases. Fig. \ref{fig_convergencePlots} shows convergence of compliance for the 2D benchmark example with the thin cutter.

\begin{table} [ht!]
	\centering
	\caption{Summary of cantilever beam results in 2D.}
	\tabulinesep=1mm
	\begin{tabu}[t!]{lcccc}
		\hline \hline 
		Tool direction &$\dfrac{V_{\Gamma_{\unc}}}{V_{\Omega_0}}$ &
		$\dfrac{\varphi_{\con}}{\varphi_{\unc}}$ & $w_{\acc}$ & $\lambda$\\
		\hline
		$(+1,~0)$   & 0.32 & 4.2 & 0.5 & 0.05 \\
		$(-1,~0)$   & 0.32 & 2.4 & 0.5 & 0.05 \\
		$(\pm1,~0)$ & 0.16 & 1.3 & 0.5 & 0.05 \\
		$(+1,+1)$   & 0.40 & 3.7 & 0.5 & 0.05 \\
		\hline
	\end{tabu}
	\label{tab_beam2Dresults}
%
\vspace{1.0em}
	\centering
	\caption{Summary of cantilever beam results in 3D.}
	\tabulinesep=1mm
	\begin{tabu}[t!]{lcccc}
		\hline \hline 
		Tool direction &$\dfrac{V_{\Gamma_{\unc}}}{V_{\Omega_0}}$ &
		$\dfrac{\varphi_{\con}}{\varphi_{\unc}}$ & $w_{\acc}$ & $\lambda$\\
		\hline
		$(+1,~0,~0)$   & 0.48 & 8.57 & 0.5 & 0.025 \\
		$(-1,~0,~0)$   & 0.53 & 8.57 & 0.5 & 0.010 \\
		$(\pm1,~0,~0)$ & 0.37 & 2.86 & 0.5 & 0.010 \\
		$(+1,+1,~0)$   & 0.54 & 2.38 & 0.5 & 0.010 \\
		\hline
	\end{tabu}
	\label{tab_beam3Dresults}
\end{table}

Fig. \ref{fig_benchmark3D} illustrates the optimized cantilever beam in 3D at
volume fraction of 0.3 with milling tools approaching from different directions
in 3D. Table \ref{tab_beam3Dresults} summarizes the results for compliance and
inaccessible volume for 3D cases. 

Table \ref{tab_timeMem} summarizes the benchmarking of computational efficiency
for FEA and IMF computations. FEA is solved using conjugate gradients and sparse
matrix-vector multiplications, while IMF is computed using convolutions via FFTs
and vectorized minimization over 4D arrays. Both computations are performed on
the GPU. According to clock times presented in Table \ref{tab_timeMem}, FEA is
the computational bottleneck as FFTs are extensively optimized for GPU computing
(using \textsf{ArrayFire} and \textsf{cuFFT}). However, since standard FFTs work
with dense matrices as opposed to FEA that can exploit sparsity, IMF computation
is the memory bottleneck on the GPU. Note that the design resolution used in TO
also dictates the tool and fixture resolutions to use the FFT-based convolution.

\begin{table*} [!t]
	\centering
	\caption{Computational time and memory of FEA vs. IMF computation using FFTs.}
	\tabulinesep=0.5mm
	\begin{tabu}[t!]{ccccccc}
		\hline \hline 
		Part Resolution & Tool Resolution & \multicolumn{2}{c@{}}{Clock Time (sec)} &
		& \multicolumn{2}{c@{}}{Memory (MB)} \\
		\cline{3-4}  \cline{6-7} 
		& &FEA & IMF& & FEA & IMF \\  
		\hline 
		\qquad 37$\times$37$\times$74 $~~\approx 1.0 \times 10^5$ &
		\qquad 142$\times$142$\times$142 $~\approx 2.9 \times 10^6$ &
		\qquad2.14 & 0.26 & & 190 & 1390 \\
		\qquad 47$\times$47$\times$93 $~~ \approx 2.3 \times 10^5$ &
		\qquad 180$\times$180$\times$178 $~\approx 5.8 \times 10^6$ &
		\qquad5.68 & 0.48 & & 386 & 2820 \\
		\qquad 54$\times$54$\times$107 $~\approx 3.1 \times 10^5$ &
		\qquad 207$\times$207$\times$205 $~\approx 8.8 \times 10^6$ &
		\qquad9.97 & 0.48 & & 586 & 4270 \\
		\qquad 59$\times$59$\times$117 $~\approx 4.1 \times 10^5$ &
		\qquad 226$\times$226$\times$224 $~\approx 1.1 \times 10^7$ &
		\qquad14.25 & 0.48 & & 764 & 5580 \\
		\qquad 63$\times$63$\times$125 $~\approx 5.0 \times 10^5$ &
		\qquad 241$\times$241$\times$241 $~\approx 1.4 \times 10^7$ &
		\qquad16.03 & 0.48 & & 938 & 6810 \\
		\hline
	\end{tabu}
	\label{tab_timeMem}
\end{table*}

\subsection{GE Bracket: 3-Axis Milling with Eye-Bolt Fixtures}

\begin{figure} [ht!]
	\centering
	\includegraphics[width=0.6\linewidth]{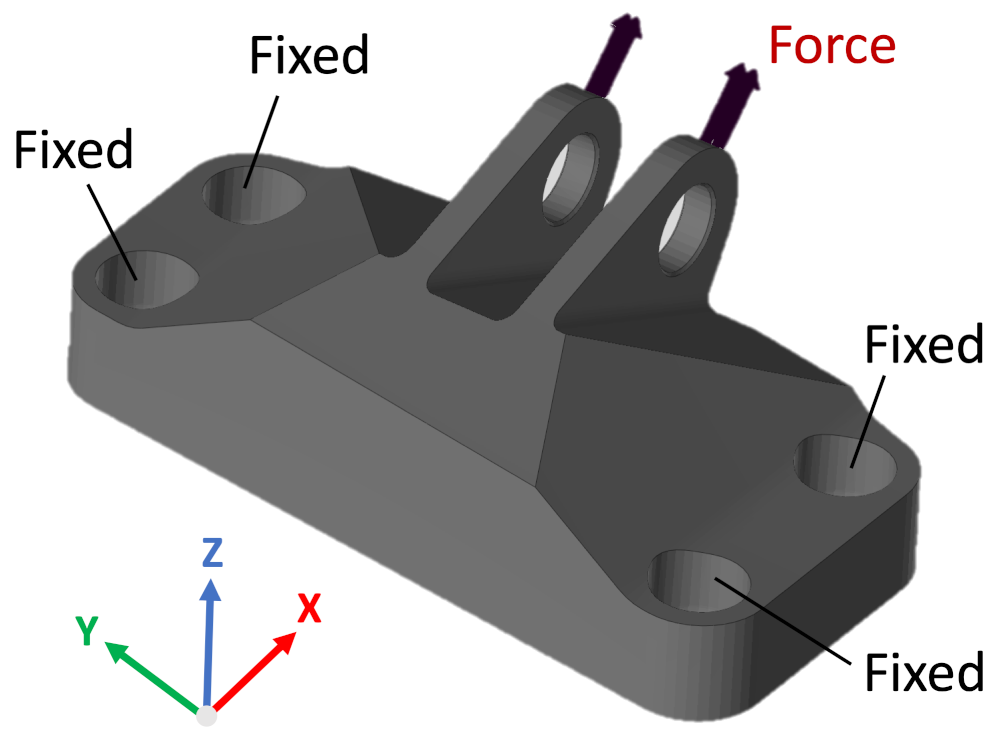}
	\caption{GE bracket loading condition.} \label{fig_GE_initial}
\end{figure}

Next, let us consider the example of GE bracket shown in Fig.
\ref{fig_GE_initial}. The material is Titanium with elastic properties of $E =
113.8$ GPa and $\nu = 0.34$. The initial design domain is discretized into
100,000 hexahedral elements.

\begin{figure} [ht!]
	\centering
	\begin{subfigure}[t]{\linewidth}
		\centering
		\includegraphics[width=0.95\linewidth]{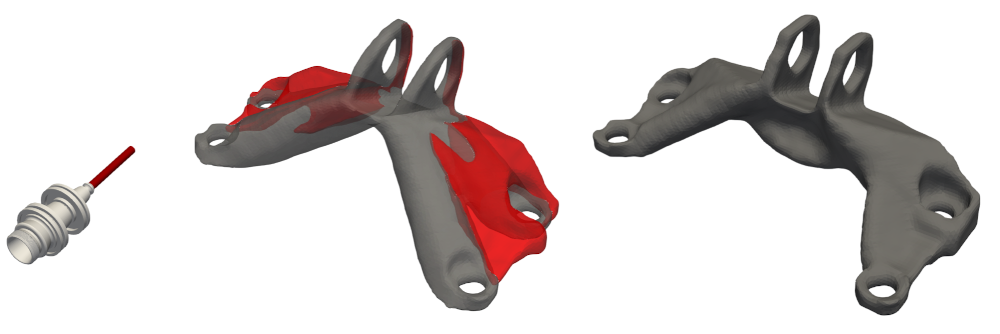}
		\caption{Tool orientation: $(+1,0,0)$}
	\end{subfigure}%
	\\
	\begin{subfigure}[t]{\linewidth}
		\centering
		\includegraphics[width=0.95\linewidth]{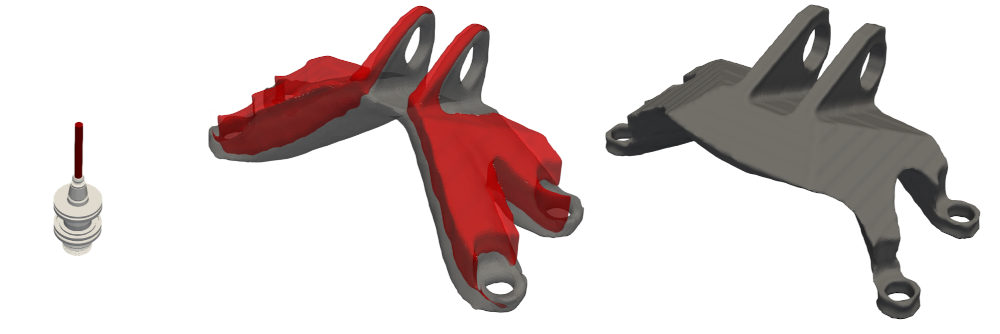}
		\caption{Tool orientation: $(0,0,+1)$}
	\end{subfigure}%
	\\
	\begin{subfigure}[t]{\linewidth}
		\centering
		\includegraphics[width=0.95\linewidth]{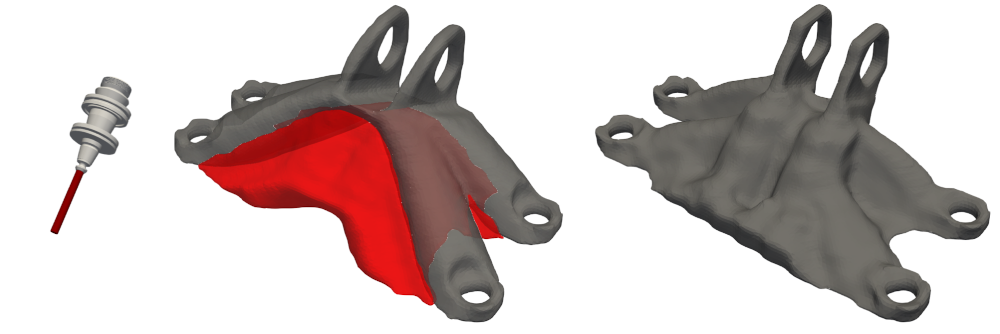}
		\caption{Tool orientation: $(-1,0,-1)$}
	\end{subfigure}%
	\\
	\begin{subfigure}[t]{\linewidth}
		\centering
		\includegraphics[width=0.95\linewidth]{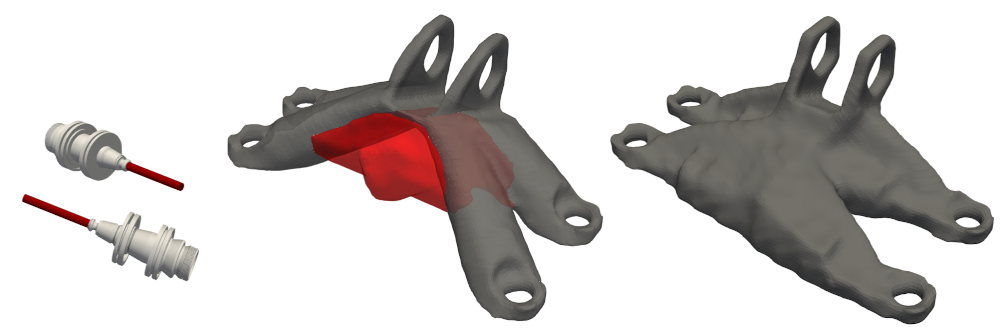}
		\caption{Tool orientation: $(0,\pm1,0)$}
	\end{subfigure}%
	\\
	\begin{subfigure}[t]{\linewidth}
		\centering
		\includegraphics[width=0.95\linewidth]{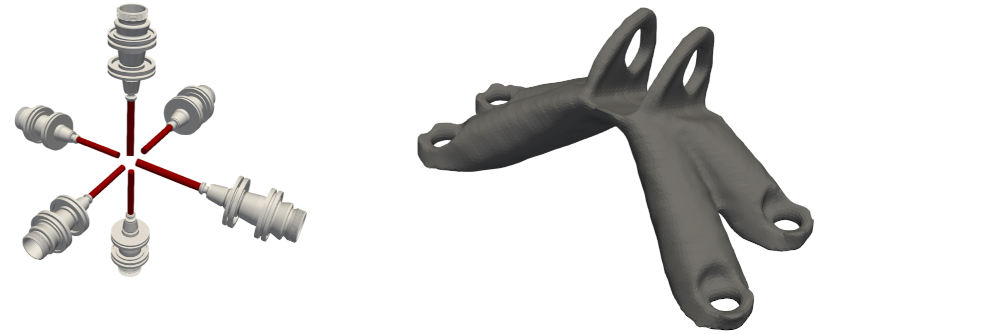}
		\caption{Tool orientations: $\begin{dcases}
			(\pm1,0,0)\\
			(0,\pm1,0) \\
			(0,0,\pm1) \\
			\end{dcases}$}
	\end{subfigure}%
	\caption{GE bracket at $0.3$ volume fraction for multi-axis milling. The unconstrained and constrained TO parts in (e) are the same.}
	\label{fig_GE_multiAxis}
\end{figure}

\begin{table}
	\centering
	\caption{Summary of GE multi-axis results.}
	\tabulinesep=1mm
	\begin{tabu}[t]{lcccc}
		\hline \hline 
		Tool direction(s) &$\dfrac{V_{\Gamma_{\unc}}}{V_{\Omega_0}}$  &
		$\dfrac{\varphi_{\con}}{\varphi_{\unc}}$ & $w_{\acc}$ & $\lambda$\\
		\hline
		$~~(+1, ~0, ~0)$   & 0.09 & 1.93 & 0.5 & 0.025 \\
		$~~(~0, \pm1, ~0)$ & 0.11 & 2.73 & 0.5 & 0.025 \\ 
		$~~(~0, ~0, +1)$   & 0.14 & 1.66 & 0.5 & 0.025 \\ 
		$~~(-1, ~0, -1)$   & 0.19 & 2.53 & 0.5 & 0.025 \\
		$\begin{dcases}
		(\pm1, ~0, ~0) \\
		(~0, \pm1, ~0) \\
		(~0, ~0, \pm1) \\
		\end{dcases}$      & 0.00 & 1.13 & 0.1 & 0.025 \\
		\hline
	\end{tabu}
	\label{tab_GEresults}
\end{table}

Fig. \ref{fig_GE_multiAxis} illustrates the optimized design at volume
fraction of $0.3$ with the milling tool approaching from different directions.
Notice that with only one or two available tool orientations, large regions of
the negative space would not be accessible for machining, hence the TO is not
allowed to remove them during the iterations. As more orientations are included,
the optimization has a larger feasible design space to explore. In every
iteration, the IMF evaluates the minimum inaccessibility over a larger range of
approach orientations. In this case, with all 6 coordinate axis-aligned
directions at hand, the result of accessibility-constrained TO begins to
resemble what one expects from unconstrained TO. Table \ref{tab_GEresults}
summarizes the results for the GE bracket example with a multi-axis milling
tool.

\begin{figure} [t!]
	\centering
	\includegraphics[width=0.8\linewidth]{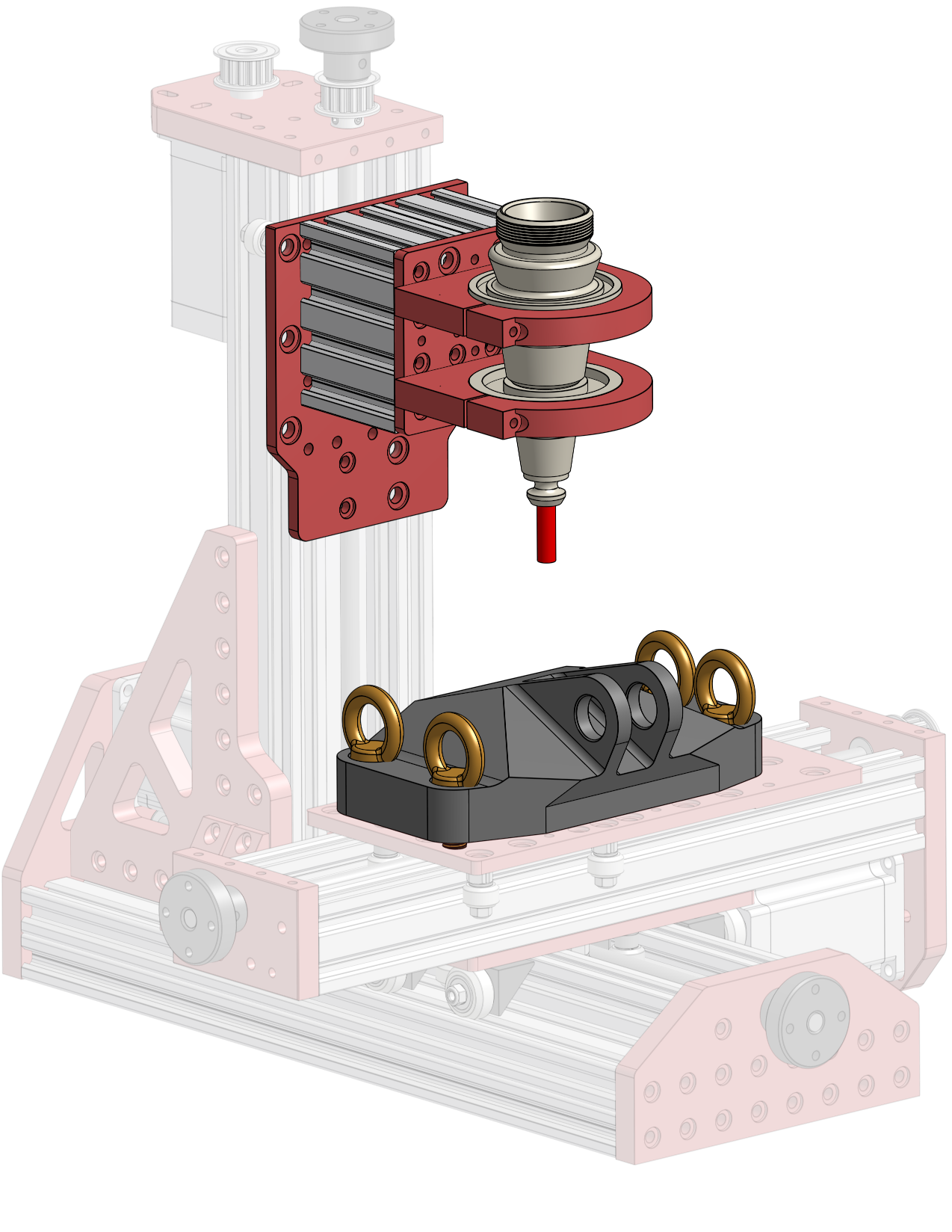}
	\caption{3-axis milling setup for the GE bracket with eye-bolt fixtures The
		tool assembly must avoid collisions with part and bolts.}
	\label{fig_CNCsetupGE}
\end{figure}
\begin{figure}[h!]
	\centering
	\begin{subfigure}[t]{\linewidth}
		\centering
		\includegraphics[width=0.8\linewidth]{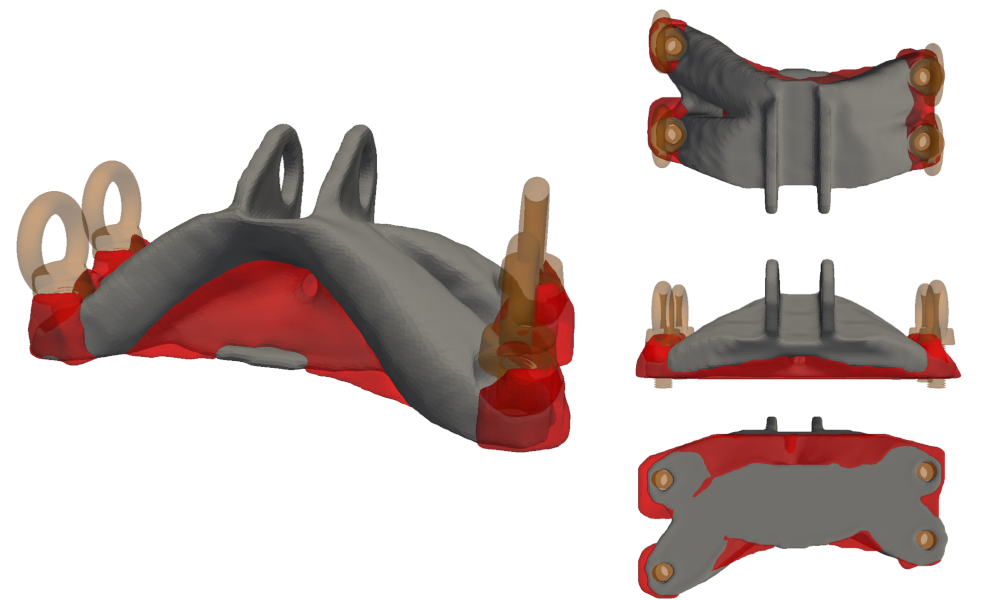}
		\caption{Optimized design via accessibility-unconstrained TO with
			inaccessible regions for machining shown in red.}
	\end{subfigure}
	\\
	\begin{subfigure}[t]{\linewidth}
		\centering
		\includegraphics[width=0.8\linewidth]{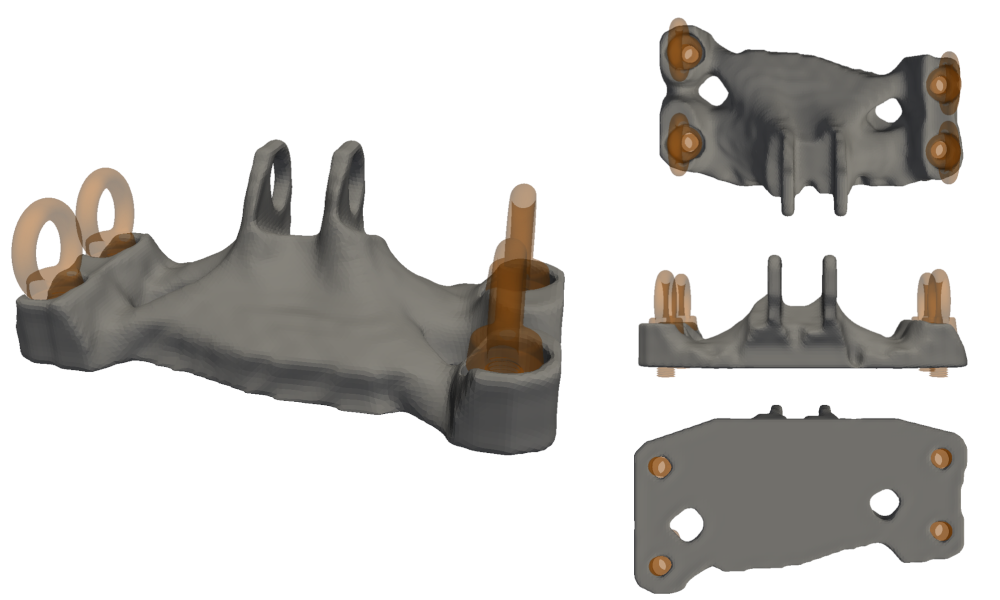}
		\caption{Optimized design via accessibility-constrained TO, machinable
			with specified tool and fixtures.}
	\end{subfigure}
	\caption{GE bracket at $0.5$ volume fraction for 3-axis milling machine with
		fixtures.}
	\label{fig_GE_CNCresults}
\end{figure}

As was discussed in Section \ref{sec_method}, it is also possible to consider
more geometrically complex tools while including fixturing devices in
accessibility analysis. For instance, consider the 3-axis CNC milling setup of
Fig. \ref{fig_CNCsetupGE} where the tool assembly consists of the cutter, tool
holder, and the clamping device. Fixturing includes 4 eye bolts that attach the
bracket to the base plate via 4 holes that are also supposed to be retained
throughout TO. Fig. \ref{fig_GE_CNCresults} (a) shows the optimized design at
a volume fraction of $0.5$ alongside the non-machinable regions (in red). Fig.
\ref{fig_GE_CNCresults} (b) shows the optimized design at the same volume
fraction with $w_{\acc} = 0.5$ and $\lambda = 0.025$. In the presence of
accessibility constraints, To takes a significantly different path in the design
space to find the stiffest possible design (in a locally optimal sense) that can
be manufactured with the given tool without colliding with the part or fixtures.

\subsection{Quadcopter: 5-Axis Milling with 3-Point Grabber}

\begin{figure} [h!]
	\centering
	\includegraphics[width=0.9\linewidth]{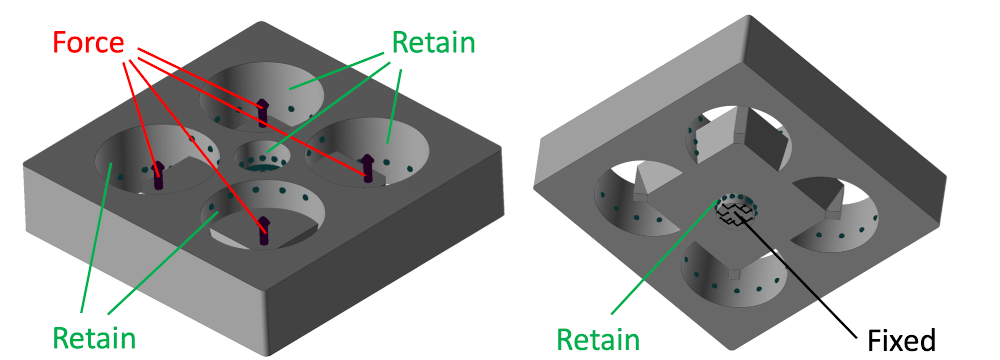}
	\caption{Quad-copter loading condition.} \label{fig_copterBC}
\end{figure}

In this section, let us consider the design of a quadcopter under hovering
loading condition as shown in Fig. \ref{fig_copterBC}. The pocket at the center
must be retained to mount battery and electronic boards. The material is
Aluminum with $E = 70$ GPa and $\nu=0.33$. Part resolution is about 300,000
voxels and target volume fraction is $0.20$.

\begin{figure}[ht!]
	\centering
	\begin{subfigure}[ht!]{\linewidth}
		\centering
		\includegraphics[width=0.85\linewidth]{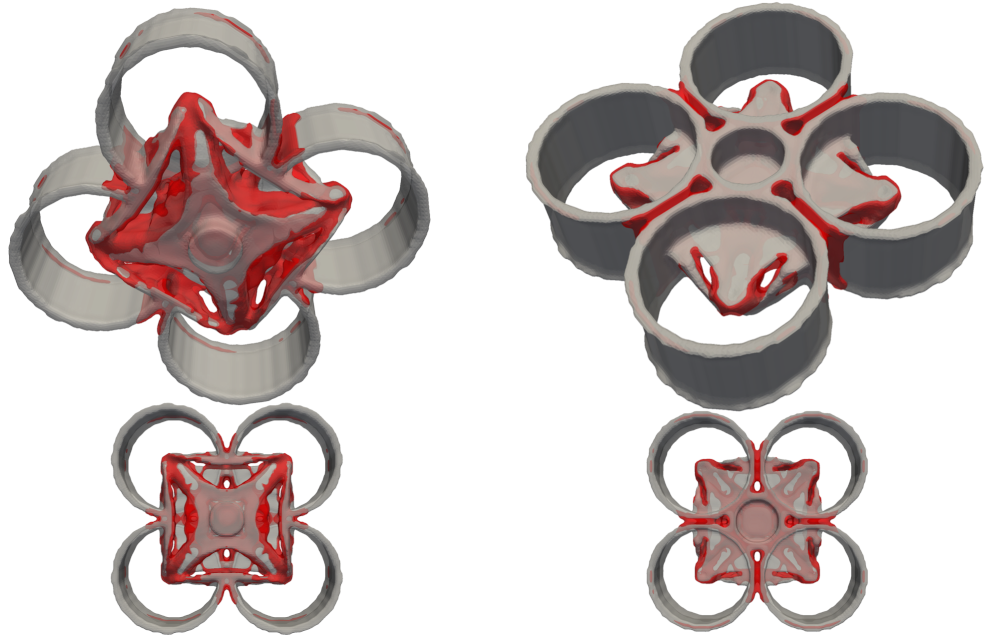}
		\caption{Optimized design via accessibility-unconstrained TO with
			inaccessible regions for machining shown in red.}
	\end{subfigure}%
	\vspace{1.0em}
	\begin{subfigure}[ht!]{\linewidth}
		\centering
		\includegraphics[width=0.85\linewidth]{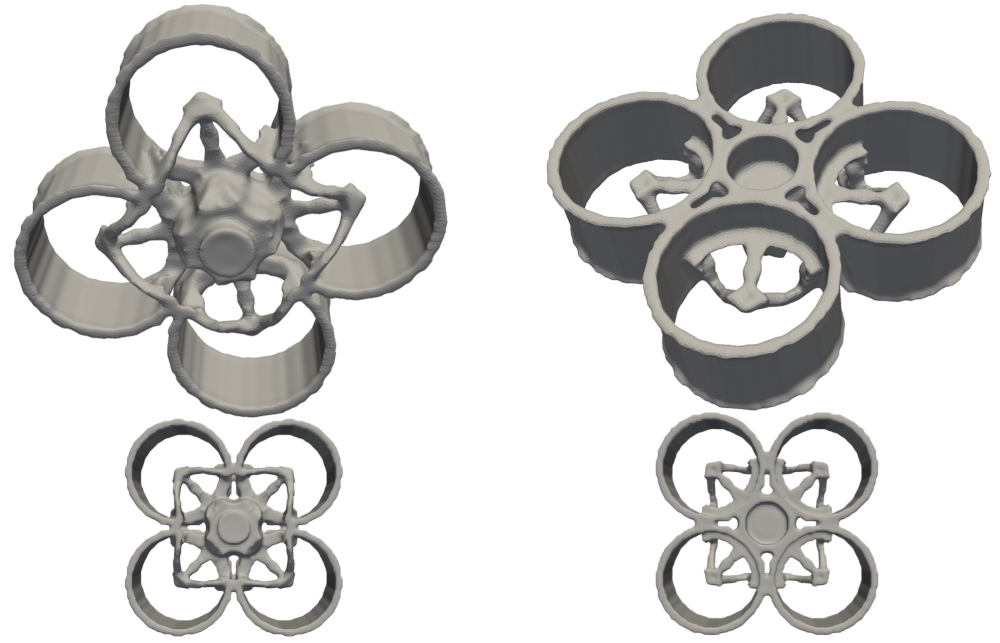}
		\caption{Optimized design via accessibility-constrained TO, machinable
			with specified tool and fixtures.}
	\end{subfigure}
	\caption{Quadcopter at $0.2$ volume fraction for 5-axis milling machine with
		3-point grabber fixture and 10 sampled orientations.}
	\label{fig_copterResults}
\end{figure}

\begin{figure} [ht!]
	\centering
	\includegraphics[width=0.85\linewidth]{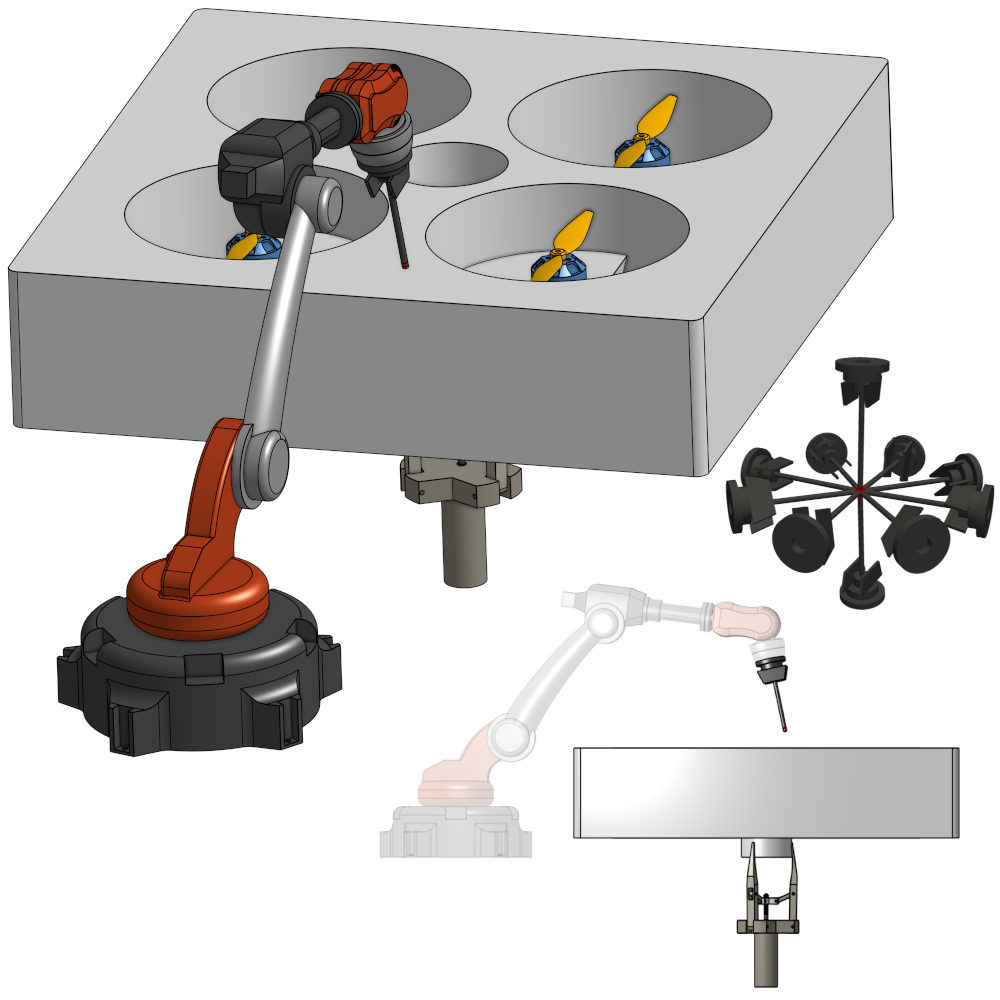}
	\caption{5-axis milling setup for quadcopter design with 10 sampled
		orientations.} \label{fig_5AxisSetup}
\end{figure}

We aim to manufacture the optimized design with the 5-axis milling robot arm as
illustrated in Fig. \ref{fig_5AxisSetup} where the raw stock is held by a
3-point grabber fixture. Figure \ref{fig_copterResults} (a) shows the optimized
design with no accessibility constraint imposed. The secluded volume ratio in
this case is ${V_{{\Gamma}_{\unc}}}/{V_{\Omega_0}} = 0.071$. Fig.
\ref{fig_copterResults} (b) shows the optimized design with accessibility
constraint using $w_{\acc} = 0.5$ and $\lambda = 0.025$, in which the secluded
volume is zeroed out.

\begin{figure} [ht!]
	\centering
	\includegraphics[width=\linewidth]{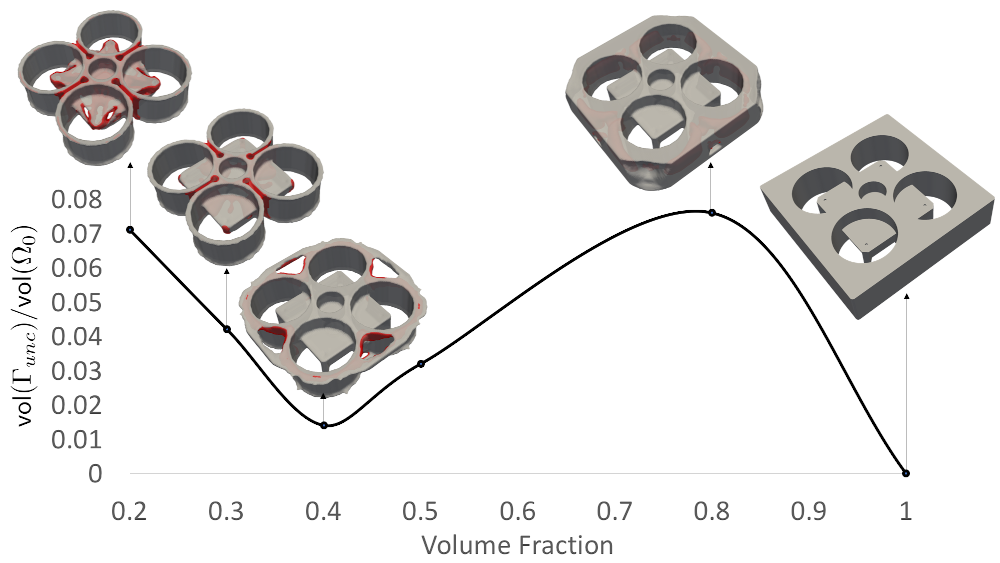}
	\caption{Inaccessible volume at different volume fractions for unconstrained
		quadcopter design.} \label{fig_InaccGraph}
\end{figure}

\begin{figure} [ht!]
	\centering
	\includegraphics[width=\linewidth]{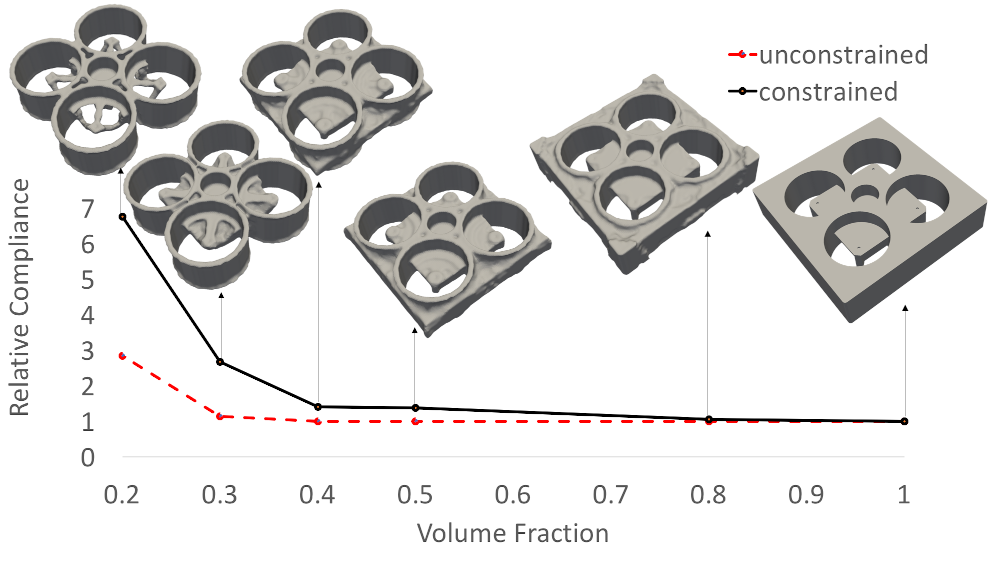}
	\caption{Relative compliance at different volume fractions for the quadcopter
		example.} \label{fig_JGraph}
\end{figure}

Fig. \ref{fig_InaccGraph} shows volume of inaccessible regions at different
volume fractions of optimized designs without considering the accessibility
constraint. Fig. \ref{fig_JGraph} shows the Pareto fronts of relative
compliance and volume fraction for both unconstrained and constrained designs.
At volume fraction of $0.20$, the accessibility constraint results in an increase
by a factor of 2.45 in the optimized compliance, i.e., $\dfrac{\varphi_{\con}}{\varphi_{\unc}}
= 2.45$ to prevent inaccessible regions.

\subsection{Support Bracket: 5-Axis Milling with Vise Fixture}

\begin{figure} [b!]
	\centering
	\includegraphics[width=\linewidth]{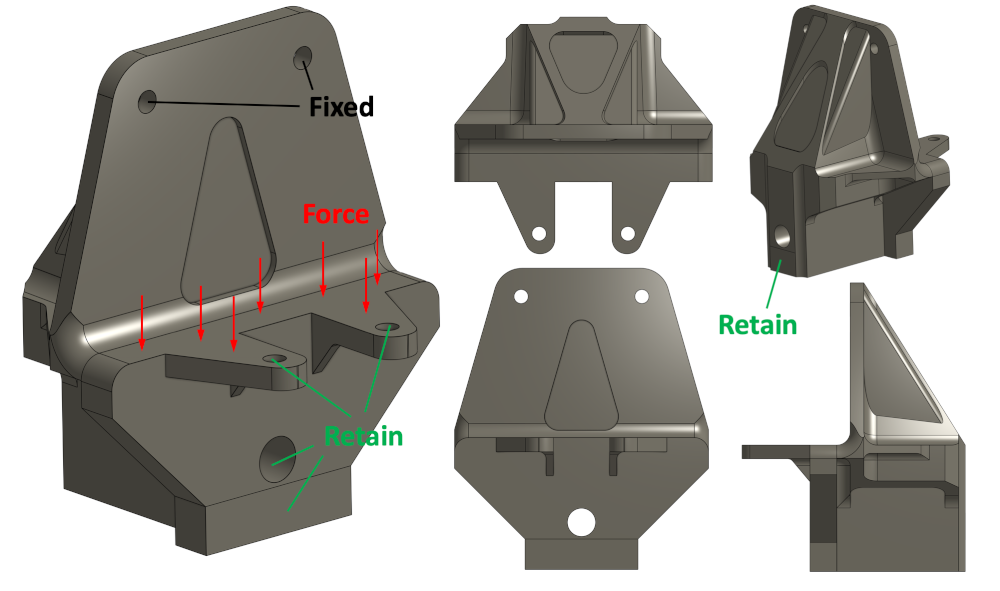}
	\caption{Support bracket geometry and loading condition.}
	\label{fig_suppLoadingp}
	%
	\centering
	\includegraphics[width=\linewidth]{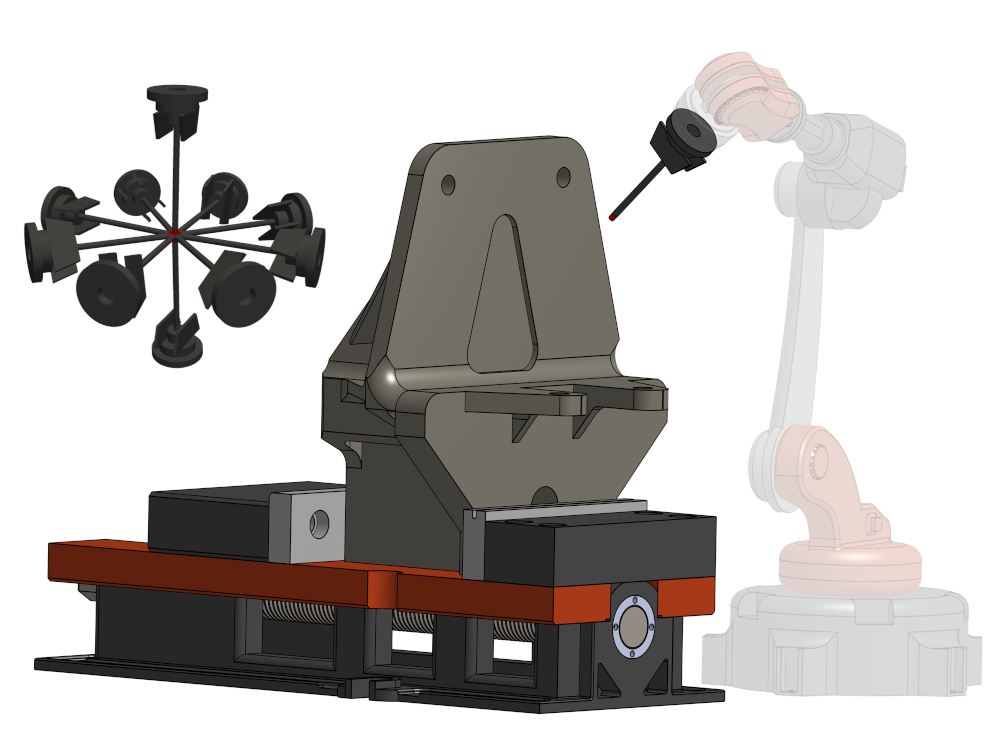}
	\caption{Support bracket setup with 5-axis milling and vise fixture. The
		highlighted components are used in accessibility analysis. The robot is
		abstracted by its DOF for orientation sampling.}
	\label{fig_supportBracketSetup}
\end{figure}

Finally, let us consider the support bracket of Fig. \ref{fig_suppLoadingp}. The
material properties are those of Stainless Steel with $E = 270$ GPa and
$\nu=0.3$. The underlying discretization is about 200,000 hexahedral finite
elements. The target volume fraction is $0.3$. The optimized design is supposed
to be fabricated with the 5-axis milling robot arm and vise fixture as
illustrated in Fig. \ref{fig_supportBracketSetup}. The highlighted components
are modeled for computing IMF and rotation space is sampled at 10 orientations
for the cutting tool. The discretized assembly of part and fixture used in
convolution has a resolution of 196$\times$149$\times$97 $\approx 1.8 \times
10^6$.

\begin{figure}[t!]
	\centering
	\begin{subfigure}[t]{\linewidth}
		\centering
		\includegraphics[width=\linewidth]{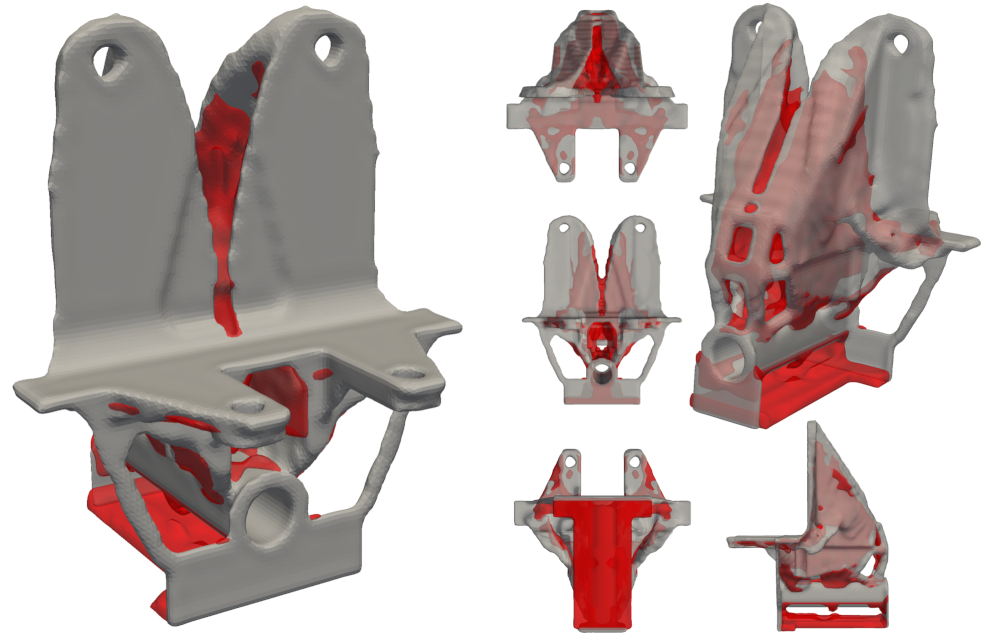}
		\caption{Optimized design via accessibility-unconstrained TO with
			inaccessible regions for machining shown in red.}
	\end{subfigure}
	\\
	\begin{subfigure}[t]{\linewidth}
		\centering
		\includegraphics[width=\linewidth]{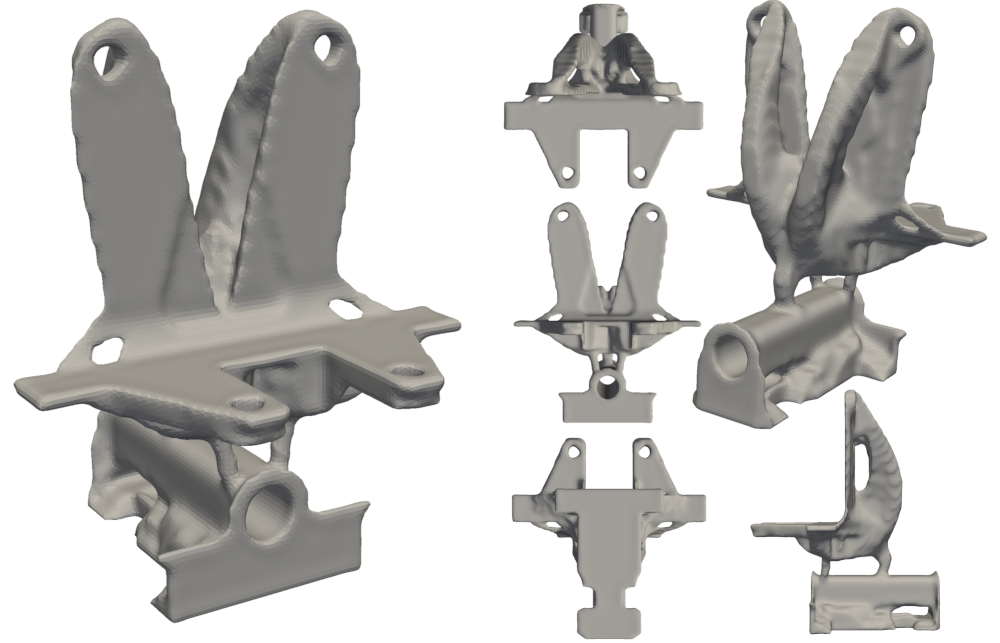}
		\caption{Optimized design via accessibility-constrained TO, machinable
			with specified tool and fixtures.}
	\end{subfigure}%
	\caption{Support bracket at $0.3$ volume fraction for 5-axis milling machine with vise fixture and 10 sampled orientations.}
	\label{fig_suppBracketResults}
\end{figure}

Fig. \ref{fig_suppBracketResults} illustrates the optimized designs for
unconstrained and constrained cases. The secluded volume
${V_{{\Gamma}_{\unc}}}$, i.e., volume of inaccessible regions outside the
optimized design for the accessibility-unconstrained TO is about $0.18~
{V_{\Omega_0}}$, which is significant. To impose the accessibility constraint,
we begin with $w_{\acc} = 0.1$ and gradually increase it to $0.5$. In this
example, the compliance of constrained and unconstrained designs are less than
$\% 1$ different. In other words, our approach produces a machinable design with
a negligible compromise in compliance.

\subsubsection{Post-Processing: Machining Process Planning}

\begin{figure*} [h!]
	\centering
	\includegraphics[width=0.85\linewidth]{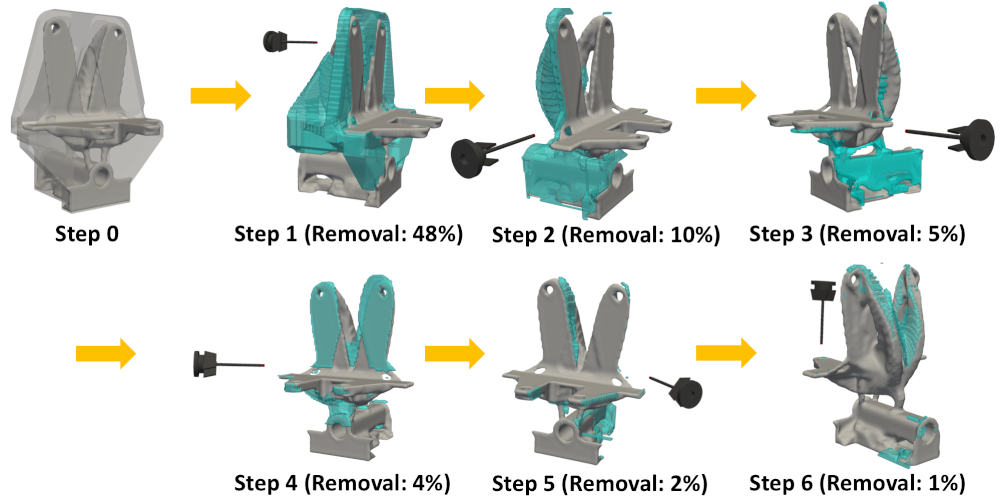}
	\caption{A 6-step machining plan based on maximum removable volume at every step. } \label{fig_supportBracketPlan}
\end{figure*}

It is important to note that the proposed TO framework {\it guarantees the
	existence of a machining process plan} with a given set of tool assemblies,
orientations, and fixtures. Once TO comes up with a design, we employ a
machining process planner to find a sequence of steps with which the negative
space can be entirely removed in as few steps as possible. The simplest
algorithm is based on a greedy criterion in terms of the maximal removable
volumes. Starting from the initial design domain, at each step we select the
oriented tool that can machine the largest volume compared to the others, and
use it to remove the subset of the negative space that is accessible to this
tool at the specified orientation. We repeat this process until the entire
negative space is removed. Figure \ref{fig_supportBracketPlan} illustrates the
6-step machining process plan to produce the accessibility-constrained optimized
design at $0.3$ volume fraction, starting from the initial design domain.

More sophisticated combinatorial optimization algorithm and practical cost
functions can be used to decide the optimal sequence of actions to be carried
out by the given tools at the available orientations
\cite{Nelaturi2015automatic,Behandish2018automated}. In any case, having the
upfront guarantee for manufacturability (i.e., existence of a process plan) and
the proper set of tools and orientations is a significant assurance before
proceeding to the costly task of process planning.

\section{Conclusion}

We presented a general approach to incorporating accessibility constraints for
multi-axis machining into existing topology optimization (TO) methods for parts,
tools, and fixtures of arbitrary shapes. We introduced the \emph{inaccessibility
measure field} (IMF) as a continuous field over the design domain to quantify
the inaccessibility of different points in the part's negative space with
respect to the geometry of intermediate design, tool assembly, fixtures, and
available orientations. The IMF is expressed via well-established mathematical
formalisms developed in spatial planning in terms of convolutions in
configuration space. We project the convolutions back to the shape space by
minimization over different candidate sharp points (on the cutter boundary) and
sampled orientations, which guarantees a conservative approximation (i.e.,
over-estimation) and computational flexibility to balance the accuracy against
time/memory budget. The methodology is fairly general and does not rely on
artificial assumptions on geometric complexity of part, tool, or fixtures.

We extended the standard TO formulation to incorporate multi-axis machining
constraints in order to reduce the discrepancy between as-designed and
as-manufactured models. Specifically, we implemented the proposed algorithm as a
density-based approach in both 2D and 3D. The effectiveness of the method was
demonstrated through benchmark and realistic examples in 2D and 3D.

Our formulation of accessibility constraint based on IMF guarantees that every
point in the design's negative space can be visited by at least one point on the
cutter in at least one orientation without incurring a collision between the
tool assembly, part, and fixtures. However, it is conceivable to encounter
situations where a collision-free configuration is not accessible {\it in a
	continuous motion} from the rest configuration, i.e., the collision-free
$\conf-$space is not path-connected. While this is limited to rare conditions
(e.g., a small tool inside a large part cavity), the IMF can be corrected by
artificially extending tool geometry via infinite half-spaces outside the
bounding box (along the approach orientation) and correcting the convolution
accordingly, or by invoking a motion planner such as Open Motion Planning
Library (OMPL) \cite{Sucan2012open}. The latter might be overkill for
intermediate iterations within TO, but can be called for post-process
validation.

Another limitation of our formulation is that accessibility and
collision-avoidance do not fully capture machining constraints. In practice,
different cutters are used for various machining operation such as milling
slots, pockets, edges, large flat surfaces, and freeform shapes. Tolerance specs and surface quality are critical when machining functional interfaces. It is
also important to consider physical constraints pertaining to the mechanics of
the cutting process. While there are studies on these issues in the context of
post-process manufacturability analysis and process planning, incorporating them
in the TO design cycles is costly and requires further research.

Future work includes investigating the extension of the method to design for
hybrid (i.e., interleaved AM/SM) sequences by developing IMF-like measures for
printability and accessibility, over-deposition (or support structure) versus
under-cut tradeoffs, and so forth. 

\section*{Acknowledgement} This research was developed with funding from the
Defense Advanced Research Projects Agency (DARPA). The views, opinions and/or
findings expressed are those of the authors and should not be interpreted as
representing the official views or policies of the Department of Defense or U.S.
Government.

\bibliographystyle{elsarticle-num} 
\bibliography{topOptAcc}

\appendix
\section{Morphology via Convolutions} \label{app_conv}

Here we explain in intuitive terms how the morphological concepts of
$\conf-$space obstacles and sweeps, used in Section \ref{sec_morph} to define
accessible regions, can be defined implicitly and computed by convolutions.

First, let us consider the first convolution in \eq{eq_conv_D} that implicitly
defines the set of colliding translations $D \subseteq \R^3$ for a fixed
rotation $R \in \SO{3}$. This convolution is defined as an integral over the
shape space (i.e., where the design resides) with a free variable in the
translation space:
\begin{equation}
	(\indic_{O} \ast \tilde{\indic}_{RT}) (\bt) = \int_{\R^3} \indic_{O}(\bx)
	\tilde{\indic}_{RT}(\bt - \bx) ~dv[\bx], \label{eq_conv_1}
\end{equation}
The integral is nonzero iff the integrand becomes nonzero over a region of
nonzero volume, meaning that both indicator functions, one of which is shifted
to apply a relative translation $\bt \in \R^3$, are nonzero. This implies a
volumetric interference between $O$ and $(R, \bt)T$, noting that $\indic_{(R,
	\bt)T}(\bx) = \indic_{RT + \bt}(\bx) = \indic_{RT}(\bx - \bt) = \indic_{-RT}(\bt
- \bx)$. The reflection $RT \to -RT$ after rotation is to flip the shifted
argument $(\bx - \bt) \to (\bt - \bx)$ to match the standard definition for
convolutions. $dv[\bx] = dx^{}_1 dx^{}_2 dx^{}_3$ for $\bx = (x_1, x_2, x_3)$ is
the volume element in the shape space.

The second convolution in \eq{eq_conv_A} that implicitly defines the sweep of
cutter along the collision-free motion can be expressed as a similar integral,
this time over the translation space, with a free variable over the shape space:
\begin{equation}
	(\neg \indic_{D} \ast {\indic}_{RK}) (\bx) = \int_{\R^3} \neg \indic_{D}(\bt)
	{\indic}_{RK}(\bx - \bt) ~dv[\bt], \label{eq_conv_2}
\end{equation}
noting that $\indic_{D^c}(\bt) = \neg \indic_D(\bt)$ is nonzero for
collision-free translations $\bt \in D^c$ and ${\indic}_{(R, \bt)K}(\bx) =
{\indic}_{RK + \bt}(\bx) = {\indic}_{RK}(\bx - \bt)$ is nonzero if the query
point $\bx \in \R^3$ is visited by the transformed cutter $(R, \bt)K$,
respectively. The convolution is nonzero if the query point is inside the sweep
$A(O, T, K)$, and the converse is true if the motion has no singularities,%
\footnote{If the motion has lower-dimensional features, they will be regularized
	as the volumetric integral cannot capture sweeps along curves or surfaces
	\cite{Behandish2017analytic}.}
as expected under quite general conditions. $dv[\bt] = dt^{}_1 dt^{}_2 dt^{}_3$
for $\bt = (t_1, t_2, t_3)$ is the volume element, this time in the translation
space.

Note that we use the $\sign$ functions in \eq{eq_conv_D} and \eq{eq_conv_A}
because indicator functions are not closed under convolutions
\cite{Behandish2017analytic}. However, for TO penalization, we do not need
binary indicator functions as we do for morphological applications. The output
of \eq{eq_conv_1} can be directly used not only to {\it classify} configurations
against the $\conf-$obstacle as in/out (as in indicator functions) but also to
{\it quantify} them in terms of a continuous measure of collision. We need is to
project this measure back to the shape space using the proper operator that
combines the collision measure for different configurations that bring different
candidate sharp points to the same point in the shape space.

Intuitively, the output of \eq{eq_conv_2} quantifies accessibility by a
summation (i.e., implicit set union) over different ways the sweep might pass
through the query point, i.e., different ways by which the sharp points can
touch the query point for machining. However, it does {\it not} quantify
inaccessibility, as it returns zero for all inaccessible points. An alternative
approach for implicitization of set union/intersection is via max/min
operations. In this paper, we apply a minimum (i.e., implicit intersection) to
the output of \eq{eq_conv_1} over different choices of sharp points, instead of using \eq{eq_conv_2}, to quantify
inaccessibility, which led to the definition of IMF in \eq{eq_imf_2} of Section \ref{sec_single}.

\section{Complexity Analysis} \label{app_comp}

For a voxelized representation of the design (as used in TO), fixtures, and tool
assembly, the convolution can be computed rapidly using FFTs
\cite{Kavraki1995computation}, and parallelized on multi-core CPUs/GPUs.  Using
the same Cartesian grid for shape voxelization and translational motion
digitization with a grid size of $n_G \gg 1$, each convolution takes two forward
FFTs, a pointwise multiplication in frequency domain, and one invewrse FFT with
time complexity of $2 O(n^{}_G \log n^{}_G) + O(n^{}_G) + O(n^{}_G \log n^{}_G)
= O(n^{}_G \log n^{}_G)$.

For $n^{}_R \geq 1$ sampled rotations, it takes $O(n^{}_R n^{}_G \log n^{}_G)$
time to compute all convolution fields, stacked together to form a single group
convolution field in the $\conf-$space.

For $n^{}_K \geq 1$ sharp points, the convolution field for each rotation is
queried at different translations $\bt = (\bx - R\bk)$, i.e., within a
$\conf-$space neighborhood $\mathfrak{N}(\bx; R) := (\bx - RK)$ that resembles a
reflected and rotated image of the cutter. It takes and additional $O(n^{}_K)$
steps to perform these queries for every oriented tool and compute the minimum,
without the need for re-computing the convolution for different shifts. Hence
the total time complexity for computing IMF for a single tool assembly is
$O(n^{}_R n^{}_G \log n^{}_G) + O(n^{}_R n^{}_K) = O(n^{}_R (n^{}_G \log n^{}_G
+ n^{}_K))$.

Finally, given $n^{}_T \geq 1$ available tool assemblies, computing the IMF
takes $O(n^{}_T n^{}_R (n^{}_G \log n^{}_G + n^{}_K))$ steps. Theoretically, as
long as $n^{}_K = O(n^{}_G \log n^{}_G)$, the sequential time complexity is
$O(n^{}_T n^{}_R n^{}_G \log n^{}_G)$. We achieve near-linear speed-up by
distributing the work on multi-core CPUs/GPUs, but the precise complexity
analysis depends on the GPU architecture \cite{Ma2014analysis}. %
If we use the same resolution to resolve the cutter as the grid size for
convolution, then $n^{}_K \ll n^{}_G$ since the cutter is typically much smaller
in size than the design domain. Even if we down-sample the sharp points to one
or a few representative points on the boundary (i.e., $n^{}_K = O(1)$) the
approximate IMF captures most of the qualitative features of the exact IMF, as
we showed for simple 2D examples in \cite{Mirzendehdel2019exploring}.

Our results show that the main bottleneck in TO iterations is the FEA. The IMF
computation using FFT-based convolution on the GPU outperforms FEA by an order
of magnitude for $n^{}_G \approx 10^5$ voxels and the discrepancy grows with the
problem size (see Table \ref{tab_timeMem} of Section \ref{sec_results}). The
main challenge with computing IMF is memory, as we need to store convolution
fields for all tool assemblies and their re-sampled grid interpolation at
different orientations. The space complexity is $O(n^{}_T n^{}_R n^{}_G)$ if all
of them are retained for rapid querying, which can be prohibitive for
high-resolution parts in large-scale TO. Given a fixed memory budget of
$O(n^{}_M)$, we serialize the parallel computations at a small time cost as
follows. We store a batch of $O(n^{}_M / n^{}_G)$ convolution fields at-a-time,
each computed on the GPU as a 3D array of size $O(n^{}_G)$. We arrange their
shifted copies for different sharp points into a 4D array of size $O(n^{}_M)$.
We compute and store the partial minimum field by pointwise comparison on the
GPU, discard the 4D array, and move on to comparing the next batch with the
partial minimum until we cover all of them.

\end{document}